\documentclass{article}
\usepackage[utf8]{inputenc}
\usepackage{authblk}
\usepackage{setspace}
\usepackage[margin=1.25in]{geometry}
\usepackage{graphicx}
\usepackage{grffile}
\graphicspath{ {./assets/} }
\usepackage{subcaption}
\usepackage{amsmath}
\usepackage{amssymb}
\usepackage{mathpazo}
\usepackage{lineno}
\usepackage{booktabs}
\usepackage{multirow}
\usepackage{makecell}
\usepackage{siunitx}
\usepackage[table]{xcolor}
\usepackage{pifont}
\usepackage{wrapfig}
\usepackage{url}

\definecolor{rowgray}{gray}{0.93}
\newcommand{\cmark}{\ding{51}}
\newcommand{\xmark}{\ding{55}}
\newcommand{\worse}[1]{\textcolor{red}{$\uparrow$#1}}
\newcommand{\better}[1]{\textcolor{green!55!black}{$\downarrow$#1}}

\usepackage[style=nejm, 
citestyle=numeric-comp,
sorting=none,
backend=biber]{biblatex}
\title{\texttt{NOCE-Net}: Representation Learning for Battery Operational Context via Nested Sequence Modelling}

\author[1,*]{Arnab Bhattacharjee}
\author[2]{Wayes Tushar}
\author[2]{Tapan K. Saha}

\affil[1]{Capalo AI, Helsinki, Finland, 00100}
\affil[2]{The University of Queensland, Brisbane, Australia, 4072}
\affil[*]{Address correspondence to: be.arnab96@gmail.com}

\date{}

\begin{document}

\maketitle

\begin{abstract}
Learning effective representations from battery operational history is essential for predictive health management and control-oriented applications. Existing approaches rely on response-dependent engineered features or short early-life windows, limiting generalisability and precluding use in prospective settings where future responses are unavailable. We propose \texttt{NOCE-Net}, a deep nested sequence model that learns control-enabling representations directly from raw cyclic operational or telemetric data. Our data pipeline, \texttt{re-BatteryData}, standardises heterogeneous aging datasets into semantically equivalent Full Equivalent Cycles (FECs) defined by absolute charge throughput, enabling end-to-end learning across diverse operational regimes. \texttt{NOCE-Net} combines multi-patch transformer encoders for intra-cycle dynamics with a GRU backbone for inter-cycle state evolution, operating on approximately $700$ batteries across ten public datasets spanning over $300$ operational profiles. In zero-shot evaluation on unseen battery chemistries and usage patterns, \texttt{NOCE-Net} achieves $13.7\%$ MAPE in voltage response prediction, outperforming strong long-sequence baselines by over $43\%$ while using $40 \times$ fewer parameters. Shape and component diagnostics of the learned hidden states reveal that the GRU's principal direction of evolution achieves Spearman $|\rho| \geq 0.95$ with the capacity fade trajectory on average across all datasets with monotone degradation profiles, including zero-shot test batteries, providing evidence that the representations encode degradation dynamics transferable to downstream tasks beyond battery response modelling. Code is available in \url{https://github.com/BeArnab96/NOCE-Net}.
\end{abstract}


\section{Introduction}\label{sec:intro}

The accelerating global transition toward renewable energy has made batteries indispensable infrastructure assets. Simultaneously, critical mineral supply constraints, volatile energy markets, and evolving policy landscapes introduce uncertainty in return-on-investment pathways, motivating the need to optimally utilize batteries to jointly maximize delivered utility and life expectancy. Battery response and degradation dynamics exhibit strong path dependence on usage history and battery type \cite{Raj2020, Geslin2024}. Consequently, learning battery type-aware generalizable representations that encode changes in battery dynamics across long horizons is essential for reliable predictive optimization and degradation monitoring \cite{Sulzer2021, Attia2020}. However, learning such representations from real-world battery data remains an open challenge.

Existing approaches to battery representation learning face four interconnected bottlenecks. \textbf{First}, most rely on \emph{response or observed operation-dependent} manually engineered features—such as $dV/dQ$ curves or partial charge statistics~\cite{9520291, 9248645, Richardson2019, Lu2023, ZHOU2025126375, Zhu2022, Zhang12025}—that require observing battery responses unavailable in prospective settings. We call a differentiable forward dynamic battery model "\emph{control-enabling}" if it is capable of predicting voltage responses over arbitrary future horizons as a function of planned operational inputs. Note that a control-enabling model should implicitly account for and propagate usage-based degradation dynamics as a function of operational inputs into those future horizons, as it determines how the voltage response turns out to be. Response-dependent feature models generally fall short of this definition. \textbf{Second}, many studies~\cite{Li_2024, Severson2019, Zhang2025, 10.1145/3711896.3737372} also assume early-life conditions remain representative throughout battery life, underestimating path-dependent degradation and leading to systematically biased lifetime predictions as operating conditions evolve. \textbf{Third}, battery aging datasets suffer from heterogeneity in resolution and recorded fields, restricting most studies to small datasets with narrow operational diversity, undermining the learning of generalizable representations~\cite{Wang2024, 10310297, Nascimento2023-or, Che2023-dl}. Existing large-scale pipelines~\cite{zhang2024batteryml, 10.1145/3711896.3737372} remain limited to retrospective SOH estimation and are not designed for control-enabling representation learning. \textbf{Finally}, health models based on analytic electrochemical or equivalent circuit models require disruptive Reference Performance Tests (RPTs)~\cite{Jones2022-lf, Lanubile2024} for parametrization, necessitating downtime.\\
To address these challenges, we propose \texttt{NOCE-Net}, a nested sequence model trained to predict arbitrarily long voltage responses from operational inputs obtained from telemetric data alone, thereby learning control-enabling representations that encode long-term battery dynamics. Operational history is segmented into variable-length full equivalent cycles (FECs), each representing a semantically identical operational period (unlike conventional definitions that treat consecutive charge–discharge events as distinct cycles, ref. Appendix~\ref{app:FEC_importance}), yielding a bilevel nested context that (i) compresses the temporal dimension for computationally efficient long-horizon supervised learning, (ii) naturally couples fast intra-cycle response dynamics with slower inter-cycle degradation dynamics, and (iii) enables non-invasive modeling of latent battery dynamics, since voltage--input relationships evolve nonlinearly with degradation.

To the best of our knowledge, transferrable, task-aware, and control-enabling representation learning from arbitrarily long operational context using deep sequence models remains largely unexplored in literature. Our contributions are three-fold:
\begin{enumerate}
    \item \texttt{re-BatteryData}. We introduce \texttt{re-BatteryData}, a unified data structure that addresses heterogeneity in battery aging datasets by generating nested operational context with uniform feature dimensions and temporal resolution, along with battery metadata extraction. Unlike \cite{zhang2024batteryml}, it is explicitly designed to support end-to-end deep representation learning from raw cyclic (telemetric) aging data.
    \item Large Li-ion Battery Dataset. We curate and process lifetime operational data from approximately 700 batteries spanning $\sim$300 unique operational profiles across 10 public datasets. In contrast to existing large-scale datasets \cite{zhang2024batteryml, 10.1145/3711896.3737372}, which primarily cover lab-controlled constant-current regimes, our dataset includes highly variable profiles such as EV drive cycles \cite{Pozzato2022}, aircraft operations \cite{Bills2021}, and dynamic partial cycling with unspecified depths of discharge \cite{Wang2024, nasaRandomizedBattery}.
    \item Nested Operational Context Encoding Network (\texttt{NOCE-Net}). We propose \texttt{NOCE-Net}, a task-conditioned deep learning framework for control-enabling \footnote{Control-enabling should not be confused with controllable. It is a much simpler construct used in this paper to determine whether a forward dynamic model is even suitable for a continuous predictive control application in the context of battery operations. The controllability of battery health by controlling its operational inputs is a question of further investigation and is out of scope for this paper. Here we are simply proposing a model that possesses the characteristics to be used as a surrogate in health-aware predictive control settings, without having to make restrictive assumptions.} representation learning from nested battery operational context. We evaluate the learned representations in zero-shot and few-shot response prediction settings, achieving $13 - 14\%$ MAPE in zero-shot scenarios and a $43\%$ improvement over the strongest baseline. Ablation studies demonstrate \texttt{NOCE-Net}’s ability to capture long-range, degradation-induced changes in battery dynamics.
\end{enumerate}
Additionally, the recurrent hidden states learned from voltage supervision encode battery degradation structure without capacity labels: the principal component of the hidden state trajectory achieves near-perfect rank correlation with capacity fade across all splits including zero-shot batteries, and shape analysis confirms that residual SOH prediction errors are affine rather than structural. A per-dataset regression head trained on
30\% of in-domain batteries achieves
5-33\% MAPE on well-behaved datasets, suggesting \texttt{NOCE-Net}'s representations are a promising foundation for downstream degradation monitoring.
MAPE on SOH trajectory prediction on well-behaved datasets. These findings suggest that \texttt{NOCE-Net}'s operational context representations are a promising foundation for downstream degradation monitoring, opening pathways that we leave to future investigation.

\begin{figure*}[ht]
  \vskip 0.2in
  \begin{center}
    \centerline{\includegraphics[width=0.9\textwidth]{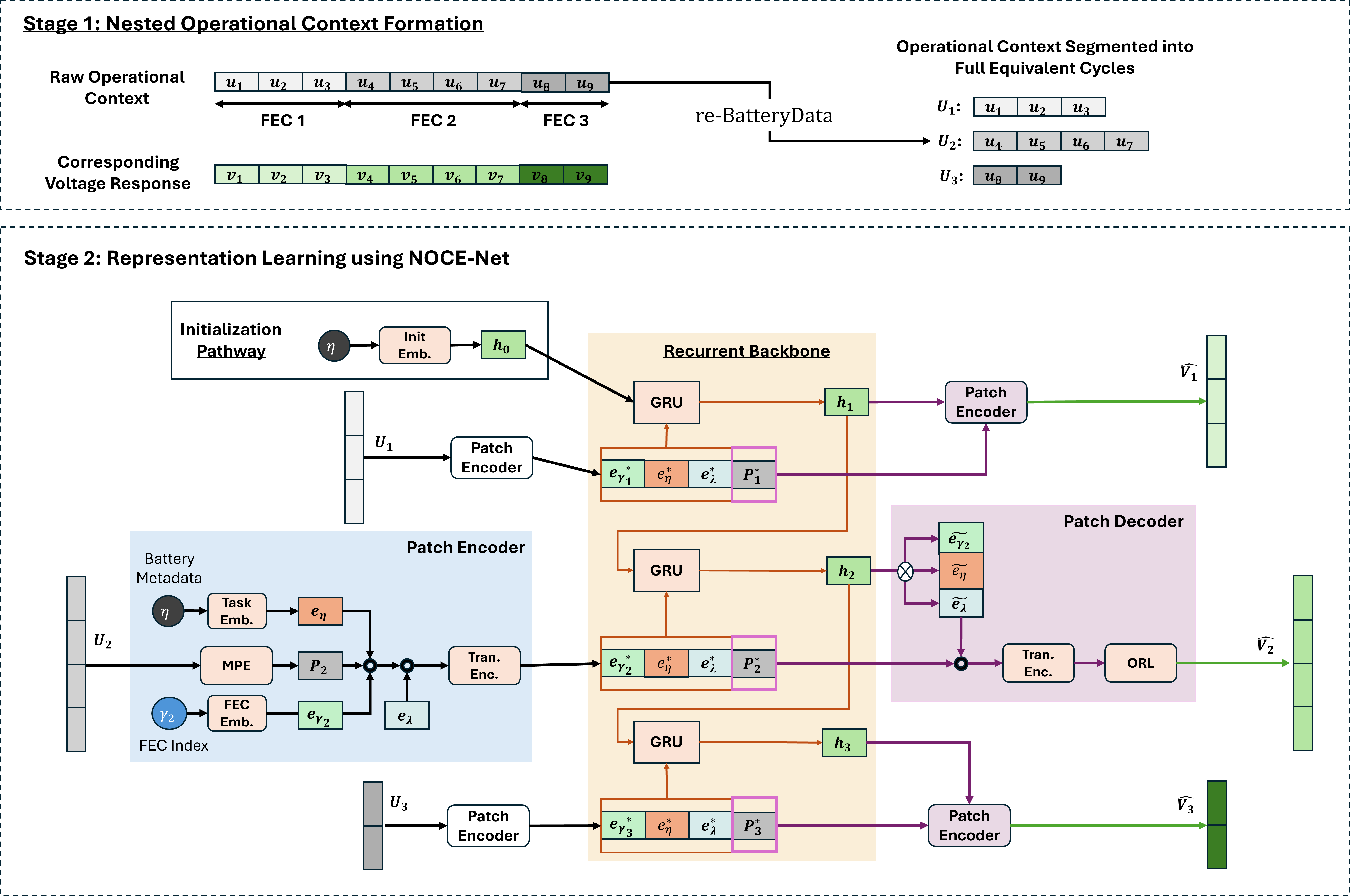}}
    \caption{
      Two-stage pipeline: \texttt{re-BatteryData} converts raw aging data into nested operational context, which is a bi-level sequence representation of battery data with the outer sequence representing equivalent cycles and the inner sequence corresponding to each equivalent cycle representing the finer inputs during each cycle. (Stage~1); \texttt{NOCE-Net} learns representations via voltage response prediction based on the nested operational context representation and learns health-indicating latents without explicit training to learn degradation (Stage~2). Abbreviations: Tran.~Enc.~=~Transformer Encoder, MPE~=~Multi-Patch Embedding, ORL~=~Output Reconstruction Layer, GRU~=~Gated Recurrent Unit.
    }
    \label{fig:overview}
  \end{center}
\end{figure*}

\section{Proposed Methodology}
As depicted in Figure 1, the battery operational context modelling pipeline operates in two stages. In the first stage, hetereogenous battery aging datasets are transformed into a uniform resolution nested operational context data using the \texttt{re-BatteryData} framework. 
In the second stage, the proposed \texttt{NOCE-Net} predicts intra-cycle voltage responses as a function of nested operational inputs using a combination of multi-patch transformer architecture for encoding intra-cycle dynamics and segment-level recurrence for modelling long-term inter-cycle dependence. This supervised learning design allows \texttt{NOCE-Net} to learn representations that preserve both local operational semantics and global battery dynamics-relevant context.

\subsection{\texttt{re-BatteryData}}\label{subsec:rebatterydata}
Given raw operational inputs $\mathbf{U} = [\mathbf{u}_1,\dots,\mathbf{u}_L]$ ($\mathbf{u}_k \in \mathbb{R}^{d_u}$, $L$ arbitrarily large) and response $\mathbf{V} \in \mathbb{R}^{L\times d_v}$ over $L$ successive timesteps, where both $\Delta L$ and $d_u$ can vary due to irregularities in data resolution and recorded features across battery datasets, \texttt{re-BatteryData} enforces four preprocessing steps:
\begin{itemize}
    \item \textbf{Uniform feature dimension}: \texttt{re-BatteryData} generates a consistent feature dimension, $d_u = 6$, consisting of continuous features - charge current ($\mathbf{I}_{chg}$), discharge current ($\mathbf{I}_{dch}$), temperature ($\mathbf{T}$), time ($\mathbf{t}$) and accumulated features - charge throughput ($\mathbf{Q}_{chg}$), discharge throughput ($\mathbf{Q}_{dch}$). Throughputs are computed by integrating the current between successive sign changes. More details can be found in section \ref{app:uniform_feats}.
    \item \textbf{Adaptive Temporal Resampling}: Continuous features are time-weighted averaged ($\mathbf{X}[\tau]=(\sum_{i \in [\tau, \tau +1)}\hat{\mathbf{X}}[i]\Delta [i])/\sum_i\Delta[i]$) and accumulated features are summed ($\mathbf{Q}[\tau]=\sum_{i \in [\tau, \tau +1)}\hat{\mathbf{Q}}[i]\Delta[i]$) to enforce a uniform temporal resolution.

\item \textbf{Battery task features} ($\boldsymbol{\eta}$): Battery specific task features ($\boldsymbol{\eta}$) containing categorical features like battery anode/cathode specifications and form factor, and continuous features like nominal capacity ($Q_{nom}$) and voltage Range ($V_{max}$, $V_{min}$), are collected for each dataset(Appendix Table~\ref{tab1}).

\item \textbf{Full equivalent cycle-based segmentation}:
Unlike existing work \cite{zhang2024batteryml, 10.1145/3711896.3737372} that rely on non-uniform charge-discharge boundaries to define a cycle, \texttt{re-BatteryData} defines a full equivalent battery cycle (FEC) as the period during which the battery undergoes an accumulated charge throughput, $Q_{acc} =\int |I| dt$, of twice its nominal capacity. This ensures that each FEC represents a semantically identical segment of operation conducted on the battery, irrespective of the operational conditions. The entire temporal record of operational context is then divided into variable-length segments based on the FEC criterion. Specifically, $\mathbf{U}_{seg}, \mathbf{V}_{seg} \gets \text{Segment}_{FEC}(\mathbf{U}), \text{Segment}_{FEC}(\mathbf{V})$, where $\mathbf{U}_{seg} = [\mathbf{U}_1, \mathbf{U}_2, \ldots, \mathbf{U}_c,\ldots, \mathbf{U}_N]$ and $\mathbf{V}_{seg} = [\mathbf{V}_1, \mathbf{V}_2, \ldots, \mathbf{V}_c,\ldots, \mathbf{V}_N]$. $N$ represents the number of FECs obtained from $L$ temporal samples, $\mathbf{U}_c = [\mathbf{u}_k, \mathbf{u}_{k+1}, \ldots, \mathbf{u}_{k+L_c}]$ is $c^{th}$ intra-cycle operational input, $\sum_{i=k}^{k+L_c} |\mathbf{I}[i]|*\delta_t \sim 2Q_{nom}$ $\forall c \in [N]$, and $\sum_{c=1}^{N} L_c = L$. More details can be found in section \ref{app:FEC_importance}
\end{itemize}
Note here that the FEC-driven boundary is a segmentation criterion, not a claim of identical degradation conditions. Within each FEC boundary, the full intra-cycle record comprising of current throughput, temperature, and time accounts for the differences in degradation pathways. Formally, \texttt{re-BatteryData}'s nested operational context formulation converts ultra-long battery telemetric historical data into a bi-level sequence representation comprising of:
\begin{enumerate}
    \item an outer sequence over full-length FECs, $c \in [N]$ and,
    \item an inner sequence containing the variable-length intra-cycle time series of operational inputs, $\mathbf{U}_c \in \mathbb{R}^{L_c \times d_u}$ nested within each FEC.
\end{enumerate}
This data representation forms the basis for all subsequent model development and analysis. 

\subsection{\texttt{NOCE-Net}}\label{subsec:nocenet}
Given a nested operational context $\mathbf{U}_{seg}$, \texttt{NOCE-Net} predicts the output response $\mathbf{V}_{seg}$, thereby learning representations that preserve long-term intra-cycle dependencies.Along with the six primary features(\ref{subsec:rebatterydata}), \texttt{NOCE-Net} additionally considers the task-specific features, $\mathbf{\eta}$ for adaptation across different battery types, and FEC-specific features,  $\gamma_c= [c, \Delta_c]$, that serve as FEC-level positional embeddings, where $\Delta_c$ represents the FEC duration.  Using these inputs, \texttt{NOCE-Net} outputs -the predicted responses $\hat{\mathbf{V}}_{seg} = [\hat{\mathbf{V}}_1, \hat{\mathbf{V}}_2,\cdots,\hat{\mathbf{V}}_c, \cdots\hat{\mathbf{V}}_N]$ where $\hat{\mathbf{V}}_c \in \mathbb{R}^{L_c \times d_v} $ and a sequence of recurrent states $\mathcal{H} = [\mathbf{h}_1, \ldots, \mathbf{h}_N] \in \mathbb{R}^{N \times d_h}$ where $\mathbf{h}_c$ encodes the cumulative operational context till FEC $c$. \texttt{NOCE-Net} comprises of three key design elements: a Patch Encoder for extracting multi-resolution representations from intra-cycle inputs, a Gated Recurrent Neural Network backbone to model inter cycle dynamics and a Patch Decoder module for reconstructing intra-cycle responses from the latent representations. Complete architectural details can be found in Appendix~\ref{app:architecture}.

\subsubsection{Patch Encoder}\label{subsubsec:pacthencoder}
For each FEC, $c \in [N]$, Patch Encoder takes the intra-cycle input, $\mathbf{U}_c \in \mathbb{R}^{L_c \times d_u}$ and applies a multi-patch embedding layer that extracts multi-resolution features across several patch lengths using multi-kernel convolutional layers. Specifically for patch sizes $\{s_1, s_2, \dots, s_K\}$, embeddings are obtained as follows:
\begin{align}
    &\mathbf{P}_c^k = \phi_{\text{conv}}^k(U_c) = \texttt{Conv1D}(U_c^\top; \text{kernel} = s_k, \text{stride} = s_k)
\end{align}
where $\mathbf{P}_c^k \in \mathbb{R}^{\left \lceil \frac{L_c}{s_k} \right\rceil  \times d_{model}}$ where $d_{model}$ is the common embedding dimension. We choose $s_k = 2^{k+a}$. The multi-kernel convolution outputs are then concatenated to form the multi-patch embedding, $\mathbf{P}_{\text{emb},c} = [\mathbf{P}_c^1, \mathbf{P}_c^2, \dots, \mathbf{P}_c^K] \in \mathbb{R}^{M_c \times d_{\text{model}}}$, where $M_c = \sum_{k=1}^{K} \left\lceil \frac{L_c}{s_k} \right\rceil$.
Additionally, battery specific task features $\mathbf{\eta}$ and FEC-specific features $\mathbf{\gamma_c}$ are embedded as follows:
\begin{align}
    & \mathbf{e}_{\eta} = \texttt{Task-Emb}(\eta) \in \mathbb{R}^{d_{model}}\\
    & \mathbf{e}_{\gamma,c} = \texttt{FEC-Emb}(\gamma_c) \in \mathbb{R}^{d_{model}}
\end{align}
\texttt{Task-Emb} and \texttt{FEC-Emb} are formed using combinations of categorical embeddings and linear projection layers, which are detailed in Appendix \ref{app:architecture}. Task embeddings $\mathbf{e}_\eta$ and FEC embeddings $\mathbf{e}_{\gamma,c}$ are concatenated with the multi-patch embedding along the temporal axis, together with a learnable global aggregator token $\mathbf{e}_{\lambda,c}$, forming $\mathbf{E}_\text{enc}^c \in \mathbb{R}^{(M_c+3)\times d_\text{model}}$. A stacked Transformer encoder~\cite{henry-etal-2020-query, xiong2020layernormalizationtransformerarchitecture, sun2023retentivenetworksuccessortransformer, shazeer2019fasttransformerdecodingwritehead} then processes this input:
\begin{equation}
    \mathbf{E}_{\text{enc},c}^* = \texttt{Trans.Enc.}(\mathbf{E}_{\text{enc},c}) = [\mathbf{P}_{c}^{*}, \mathbf{e}_{\eta}^*, \mathbf{e}_{\gamma,c}^{*}, \mathbf{e}_{\lambda,c}^{*}] \in \mathbb{R}^{(M_c + 3) \times d_{model}}
\end{equation}
where the last three tokens - $\mathbf{e}_{\lambda,c}^{*}, \mathbf{e}_{\eta}^*, \mathbf{e}_{\gamma,c}^{*}$ - aggregate unconditional as well as task and FEC conditioned cycle-level information. The Patch Encoder operates in parallel across all FECs.

\subsubsection{Gated Recurrent Network Backbone}\label{subsubsec:gru}
The aggregated intra-cycle embeddings,  $\mathbf{E}_{agg,c} = [\mathbf{e}_{\eta}^*, \mathbf{e}_{\gamma,c}^{*}, \mathbf{e}_{\lambda,c}^{*}]$ are used as inputs in a Gated Recurrent Neural Network (GRNN) to model inter-cycle dynamics. $\mathbf{E}_{agg,c}$ is first reshaped into $\tilde{H}_c = \texttt{Concat}(\mathbf{E}_{agg,c}, dim= -1) \in \mathbb{R}^{ 3d_{\text{model}}}$ by concatenating the aggregated features along the feature dimension and then input into a Gated RNN as shown:
\begin{align}
    &\mathbf{h}_c = \texttt{Gated-RNN}(\tilde{H}_c, \mathbf{h}_{c-1})\\
    & \mathbf{h}_0 = \phi_{\text{init}}(\mathbf{\eta}) = \texttt{Task.Emb.}_{\texttt{init}}(\mathbf{\eta})
\end{align}
We use Gated Recurrent Unit (GRU) as the GRNN block. Here $\mathbf{h}_c \in \mathbb{R}^{3d_{model}}$ encodes cumulative operational context to FEC $c$. The task-conditioned initialisation $\mathbf{h}_0 = \phi_\text{init}(\boldsymbol{\eta})$ enables task-aware generalization across battery types. This GRNN framework operates only on high-level fixed dimensional cycle aggregators, allowing efficient training on GPUs. It allows expensive intra-cycle transformer operations to be conducted in parallel across the whole operational context leaving a significantly smaller number of operations that need to be conducted sequentially.

\subsubsection{Patch Decoder: Voltage Response Prediction}\label{subsubsec:patchdecoder}
The patch decoder reconstructs variable-length intra-cycle voltage responses from the attended patch embeddings $\mathbf{P}_{c}^{*}$ and the inter-cycle state $\mathbf{h}_c$. First, $\mathbf{h}_c \in \mathbb{R}^{3d_{model}}$ is reshaped to $\mathbf{H}_{\text{dec},c} \in \mathbb{R}^{3 \times d_{model}}$ and is concatenated along the temporal axis to the attended multi-resolution embeddings, $ \mathbf{P}_{c}^{*}$ to form the patch decoder input $ \mathbf{E}_{\text{dec}, c} = [\mathbf{P}_{c}^{*}, \mathbf{H}_{\text{dec},c}] \in \mathbb{R}^{(M_c+3) \times d_{model}}$. A stacked transformer encoder is applied to this input as shown:
\begin{equation}
    \hat{\mathbf{Z}}_{\text{dec},c} = \texttt{Trans.Enc.}( \mathbf{E}_{\text{dec}, c}) = [\tilde{\mathbf{P}}_{c}, \tilde{\mathbf{H}}_{\text{dec},c}]
\end{equation}
where $\hat{\mathbf{Z}}_{\text{dec},c}\in \mathbb{R}^{(M_c+3) \times d_{model}}$. An output reconstruction layer (ORL) applies hierarchical transposed convolutions with residual fusion across patch scales on patch level embeddings of $\tilde{\mathbf{P}}_{c}$, to upsample to the original intra-cycle length $L_c$, followed by an MLP:
\begin{align}
     & \mathbf{Y}_c^{K} = \texttt{TransConv1D}({\tilde{(\mathbf{P}}_{c}^K})^\top , \text{kernel} = m_K, \text{stride} = m_K) \\
    & \mathbf{Y}_c^{k-1} = \texttt{TransConv1D}(({\mathbf{Y}_c^k} + {\tilde{\mathbf{P}}_{c}}^{k-1})^\top, \text{kernel} = m_k, \text{stride} = m_k) \\
    &\hat{\mathbf{V}}_c = \texttt{MLP}(\mathbf{Y}_c^1) \in \mathbb{R}^{L_c \times d_v}
\end{align}
where  $[\tilde{\mathbf{P}}_{c}^1, \tilde{\mathbf{P}}_{c}^2, \ldots, \tilde{\mathbf{P}}_{c}^K]$ is the patch level decomposition of $\tilde{\mathbf{P}}_{c}$ with $K$ patches. $m_k = \left\lceil s_k/s_{k-1} \right\rceil  $ where $s_k \forall k \in [K]$ are the Patch Encoder patch sizes. The decoder ensures that the model’s final output is temporally aligned with the intra-cycle resolution thereby enabling intra-cycle voltage response prediction across variable length FECs.\\

\section{Experimental Details}\label{sec:expdetails}
\subsection{Dataset and Training}\label{subsec:datasets}

\begin{table*}[t]
\centering
\caption{Summary of datasets used for training and evaluation, covering diverse chemistries, form factors, and operational protocols.}
\label{tab:datasets}
\resizebox{\textwidth}{!}{
\begin{tabular}{lccccccccccc}
\toprule
\textbf{Dataset Name} & \textbf{Cathode} & \textbf{Anode} & \textbf{$Q_{nom}$ (Ah)} & \textbf{$V_{max}$} & \textbf{$V_{min}$} & \textbf{Avg. FEC Len.} & \textbf{Std. Len.} & \textbf{Operating Protocol} & \textbf{DOD} & \textbf{\# Bat.} & \textbf{Split} \\ 
\midrule
ISU-ILCC\cite{https://doi.org/10.25380/iastate.22582234} & NCM & Graphite & 0.25 & 4.2 & 3.0 & 1187 & 706.7 & Variable Charge/Discharge & Partial & 155 & Train \\ 
NASA-Rand\cite{nasaRandomizedBattery} & NMC & Graphite & 2.1 & 4.2 & 3.2 & 375.4 & 61.7 & Random Discharge & Partial & 13 & Train \\ 
TJU\cite{Zhu2022} & NCA & Graphite & 3.5 & 4.2 & 2.5 & 357 & 282.15 & Constant Charge/Discharge & Full & 100 & Train \\ 
MATR\cite{Severson2019} & LFP & Graphite & 1.1 & 3.6 & 2.0 & 795.2 & 373.83 & Const. Disch./Multi-Step Charge & Full & 124 & Train \\ 
UMICH\cite{https://doi.org/10.7302/7tw1-kc35} & NMC & Graphite & 5.0 & 4.2 & 3.0 & 231 & 89.67 & Variable Charge/Discharge & Partial & 21 & Train \\ 
XJTU-Partial\cite{Wang2024} & NCM & Graphite & 2.0 & 4.2 & 2.5 & 512 & 286.79 & Constant Charge/Discharge & Partial & 16 & Train \\ 
evTOL\cite{Bills2021} & Unknown & Unknown & 3.0 & 4.2 & 2.5 & 450 & 263.75 & eVTOL Aircraft Discharge Profile & Partial & 14 & Test \\ 
NASA-PCoE\cite{SahaGoebel2007NASA} & NMC & Graphite & 2.0 & 4.2 & 2.5 & 112 & 75.68 & Constant Charge/Discharge & Full & 10 & Test \\ 
Stanford-EV\cite{Pozzato2022} & NMC & Graphite/Si & 4.85 & 4.2 & 2.5 & 311 & 65.49 & EV Charge/Discharge Profiles & Partial & 6 & Test \\ 
HUST\cite{Ma2022} & LFP & Graphite & 1.1 & 3.6 & 2.0 & 1898 & 389.87 & Variable Charge/Discharge & Full & 77 & Test \\ 
XJTU-Full\cite{Wang2024} & NCM & Graphite & 2.0 & 4.2 & 2.5 & 441 & 189.64 & Variable Charge/Discharge & Full & 39 & Test \\ 
\bottomrule
\end{tabular}
}
\vspace{-3mm}
\end{table*}
\begin{figure}[t]
    \centering
    \begin{minipage}[t]{0.24\textwidth}
        \centering
        \includegraphics[width=\textwidth]{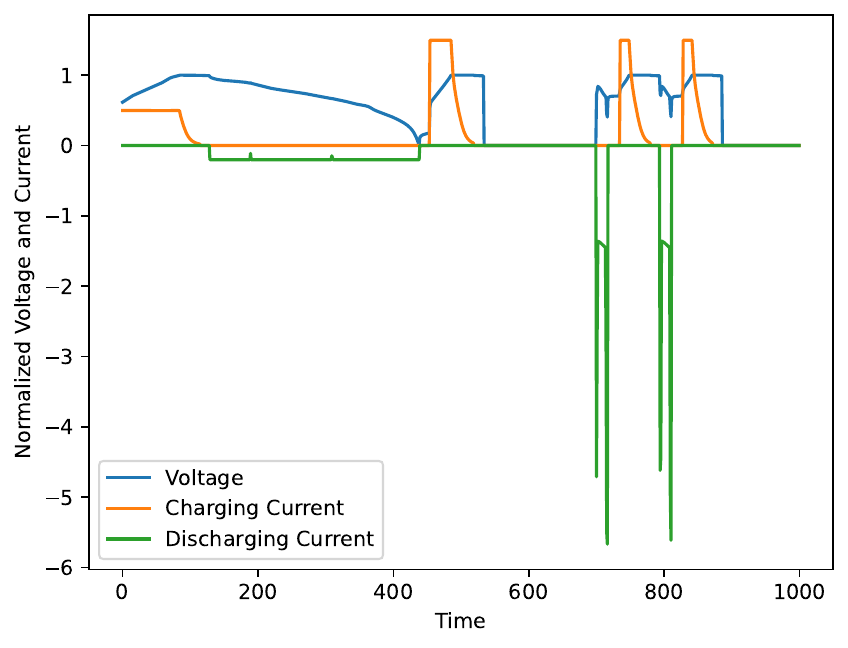}
        \subcaption{evTOL}
        \label{fig:evTOL}
    \end{minipage}
    \hfill
    \begin{minipage}[t]{0.24\textwidth}
        \centering
        \includegraphics[width=\textwidth]{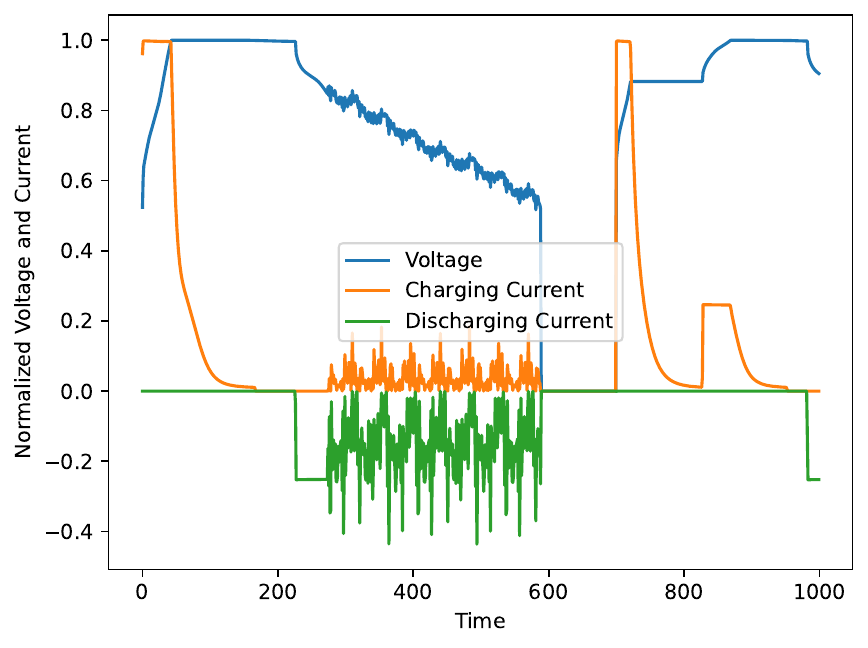}
        \subcaption{Stanford-EV}
        \label{fig:stanford}
    \end{minipage}
    \hfill
    \begin{minipage}[t]{0.24\textwidth}
        \centering
        \includegraphics[width=\textwidth]{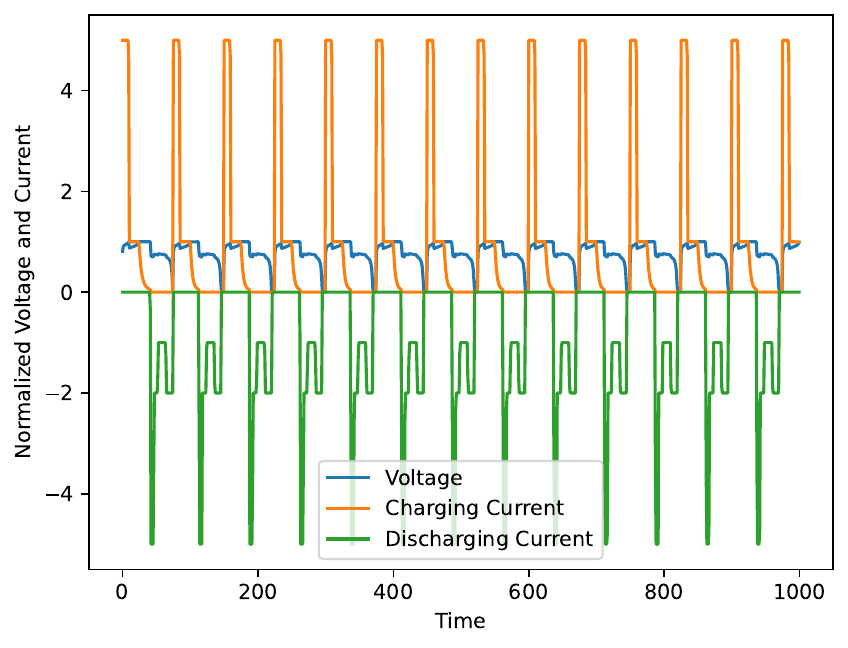}
        \subcaption{HUST}
        \label{fig:hust}
    \end{minipage}
    \hfill
    \begin{minipage}[t]{0.24\textwidth}
        \centering
        \includegraphics[width=\textwidth]{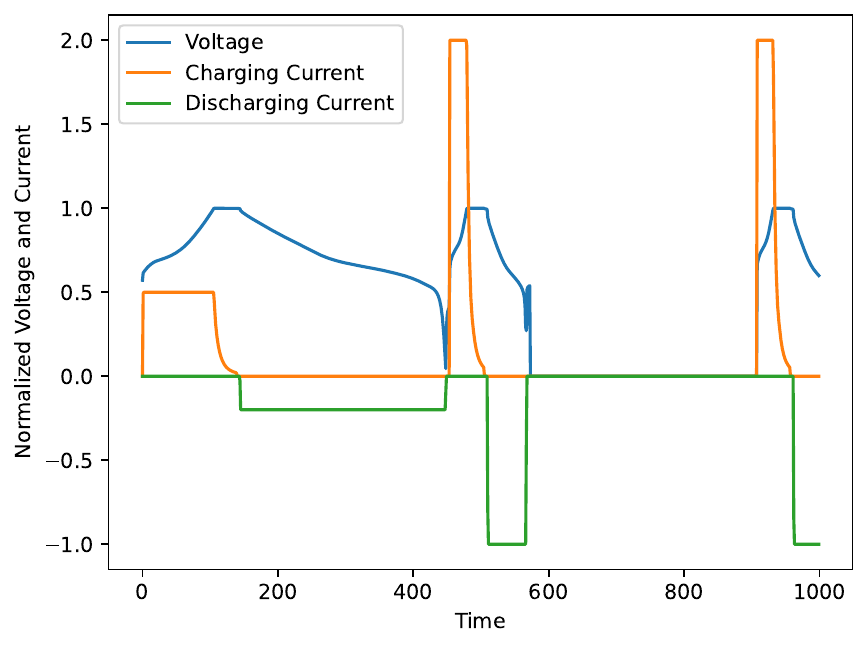}
        \subcaption{XJTU-Full}
        \label{fig:xjtu}
    \end{minipage}
    \caption{Representative operational profiles of (a) evTOL, (b) Stanford-EV, (c) HUST, and (d) XJTU-Full datasets.}
    \label{fig:all_ip_profiles}
\end{figure}

We collect a large battery aging dataset by joining 10 publicly available datasets, with 700 batteries representing around 300 different operational profiles as depicted in Table \ref{tab:datasets}. The XJTU dataset is split into XJTU-Full containing full cycling conditions and XJTU-Partial containing partial cycles. Figure \ref{fig:all_ip_profiles} shows representative voltage and current profiles for 4 such datasets. Battery datasets marked "\emph{Train}" in Table \ref{tab:datasets} - ISU-ILCC, MATR, XJTU-Partial, NASA-Rand, UMICH-Pouch, and TJU -  are used for training, validation and in-sample testing with a split of 6:2:2. The held-out in-sample test batteries are denoted by \textbf{Seen} — unseen batteries from seen chemistries. 
Additionally, test batteries are held out at the dataset level. The five datasets marked "\emph{Test}" - NASA-PCoE, Stanford-EV, HUST, XJTU-Full, evTOL - are excluded from training entirely — no battery from them is ever seen at any stage — and constitute the set, \textbf{Unseen} - the test set used to evaluate zero-shot and few-shot performance overcross-chemistry and cross-operational-regimes.

All baselines and \texttt{NOCE-Net} are trained to predict voltage responses across the lifetime of the batteries in the training set. \texttt{NOCE-Net} is trained using a composite loss function comprising of a MSE reconstruction loss, $\mathcal{L}_{MSE}$, a differential preserving loss, $\mathcal{L}_{Diff}$ and a tail penalizing loss,$\mathcal{L}_{Q}$ (Ref. Appendix \ref{app:datasets} and \ref{app:lossfn} for more details on the datasets and loss function respectively). 

\subsection{Baselines}
\texttt{NOCE-Net} is compared against TimerXL~\cite{liu2025timerxl}, AutoTimes~\cite{NEURIPS2024_dcf88cbc}, MAMBA~\cite{gu2024mamba}, and TinyTimeMixers (TTM)~\cite{ekambaram2024tiny}, covering all major architectural classes in long context time series models, using implementations from \url{https://github.com/thuml}. We evaluate three modes - M1 (nested context with binary controls), M2 (flat context) and Indep. (per cycle), of applying these long-sequence baseline models to battery data, designed to isolate the contribution of the nested operational context structure itself (ref. Table \ref{tab:mode_comparison}). Concretely, for a battery with 3 FECs, $C_1, C_2, C_3$ of lengths $L_1$, $L_2$, $L_3$ resp.,:
\begin{enumerate}
    \item \textbf{M1} treats the data bi-levelly, exactly as NOCE-Net does – each cycle is encoded separately and passed through the baseline's inter-cycle mechanism.
    \item \textbf{M2}  sees a single unbroken flattened sequence of length, $L_1 + L_2 + L_3$, and the nested boundary information is discarded entirely.
    \item \textbf{Indep.} sees $C_1, C_2, C_3$  as three completely independent sequence samples of lengths, $L_1$, $L_2$, $L_3$, with no shared state between them.
\end{enumerate}
Two binary flags in \textbf{M1} control exactly how the baseline handles the inner sequence:
\begin{itemize}
    \item \textbf{MPATCH}: whether intra-cycle inputs are embedded with a multi-patch convolutional layer (True) or a plain MLP (False). When (True), the output reconstruction layer from NOCE-Net is applied to generate the voltage response, which in the (False) case is created using an MLP head in the time dimension.
    \item \textbf{\textsc{VarSeq}} whether variable intra-cycle lengths are preserved via max-length batch padding (True) or attention-pooled to a fixed length (False).
\end{itemize}

\textbf{M1} with (\textbf{MPATCH} = True) baselines serve as targeted hierarchical comparisons on the \texttt{re-BatteryData} format without \texttt{NOCE-Net}-specific conditioning. Comparing \textbf{M1} vs. \textbf{M2} for the same architecture directly tests the value of preserving the nested structure, while \textbf{Indep.} baselines help isolate the contribution of inter-cycle memory.
Baselines use the same loss function trained for 50 epochs on an 80GB A100; \texttt{NOCE-Net} was trained on a 22.5GB L40. Several configurations were omitted due to OOM errors. Full baseline descriptions are in Appendix~\ref{app:baselines}.

\section{Results and Discussions}\label{sec:results}

\subsection{Zero-shot and Few-shot Response Forecasting}\label{sec:res_zeroandfewshot}

\begin{table*}[t]
\caption{Zero-shot and few-shot voltage prediction results. Total-Seen and Total-Unseen are macro-averages over in-sample and out-of-sample test batteries respectively. Values are means over 5 runs, (std. $5\%$-$10\%$, higher when \textsc{VarSeq}=True). Independent and TimerXL baselines omitted from few-shot due to computational overhead and consistent underperformance respectively. \textbf{Bold}~=~best, \underline{underline}~=~second best.}
\label{tab:zero-shot}
\centering
\scriptsize
\setlength{\tabcolsep}{2.5pt}
\renewcommand{\arraystretch}{1.1}
\resizebox{\textwidth}{!}{%
\begin{tabular}{llll
  S[table-format=1.4] S[table-format=2.3]
  S[table-format=1.4] S[table-format=2.3]
  S[table-format=1.4] S[table-format=2.3]
  S[table-format=1.4] S[table-format=2.3]
  S[table-format=1.4] S[table-format=2.3]
  S[table-format=1.4] S[table-format=2.3]
  S[table-format=1.4] S[table-format=2.3]
  S[table-format=1.4] S[table-format=2.3]}
\toprule
 & & & &
 \multicolumn{2}{c}{\textbf{NASA-PCoE}} &
 \multicolumn{2}{c}{\textbf{Stanford-EV}} &
 \multicolumn{2}{c}{\textbf{HUST}} &
 \multicolumn{2}{c}{\textbf{XJTU-Full}} &
 \multicolumn{2}{c}{\textbf{evTOL}} &
 \multicolumn{2}{c}{\textbf{Total-Unseen}} &
 \multicolumn{2}{c}{\textbf{Total-Seen}} &
 \multicolumn{2}{c}{\textbf{FewShot-Unseen}}\\
\cmidrule(lr){5-6}\cmidrule(lr){7-8}\cmidrule(lr){9-10}
\cmidrule(lr){11-12}\cmidrule(lr){13-14}\cmidrule(lr){15-16}
\cmidrule(lr){17-18}\cmidrule(lr){19-20}
\textbf{Method} & \textbf{VS} & \textbf{MP} & \textbf{Model}
& {MAE} & {MAPE} & {MAE} & {MAPE} & {MAE} & {MAPE}
& {MAE} & {MAPE} & {MAE} & {MAPE} & {MAE} & {MAPE}
& {MAE} & {MAPE} & {MAE} & {MAPE}\\
\midrule

\multirow{4}{*}{\textbf{M1}}
& F & F & \multirow{4}{*}{\textbf{AutoTimes}}
& 0.1782 & 31.158 & 0.1505 & 18.565 & 0.1431 & 40.420 & 0.1300 & 23.532 & 0.1852 & 27.954 & 0.1474 & 34.000 & 0.1598 & 20.709 & 0.1328 & 33.248\\
& F & T  &
& 0.1780 & 31.675 & 0.1313 & 16.829 & 0.1395 & 41.239 & 0.0765 & 14.846 & 0.1317 & 22.878 & 0.1288 & 32.745 & 0.1470 & 18.634 & 0.0702 & 33.909\\
& T  & F &
& 0.1881 & 32.664 & 0.1168 & 15.545 & \underline{0.0921} & 55.821 & 0.1215 & 21.507 & 0.1489 & 23.564 & 0.1067 & 42.870 & 0.1802 & 27.317 & 0.136 & 61.623\\
& T  & T  &
& 0.1926 & 33.367 & \underline{0.1190} & 16.592 & 0.1275 & 77.846 & 0.1783 & 24.561 & 0.1513 & 23.636 & 0.1380 & 58.161 & 0.2066 & 32.524 & 0.1243 & 57.54\\
\cmidrule(lr){2-20}

\multirow{4}{*}{\textbf{M1}}
& F & F & \multirow{4}{*}{\textbf{TTM}}
& 0.1882 & 31.679 & 0.1304 & 16.866 & 0.1298 & 38.978 & 0.1348 & 22.932 & 0.1874 & 28.474 & 0.1383 & 33.233 & 0.1576 & 20.355 & 0.1255 & 32.231\\
& F & T  &
& 0.1338 & 22.943 & 0.1704 & 21.015 & 0.1256 & \underline{31.276} & \underline{0.0467} & \underline{8.7408} & 0.1207 & 18.862 & 0.1172 & 25.365 & 0.1229 & \underline{14.633} & 0.0593 & 31.88\\
& T  & F &
& 0.1923 & 33.680 & 0.1204 & 15.923 & 0.0967 & 65.623 & 0.1557 & 28.140 & 0.1539 & 23.990 & 0.1158 & 50.109 & 0.1964 & 30.228 & 0.1547 & 56.181\\
& T  & T  &
& 0.1829 & 34.025 & 0.1193 & \underline{16.489} & 0.1300 & 76.626 & 0.1806 & 33.216 & 0.1440 & 23.552 & 0.1389 & 57.771 & 0.2089 & 33.384 & 0.1094 & 49.192\\
\cmidrule(lr){2-20}

\multirow{4}{*}{\textbf{M1}}
& F & F & \multirow{4}{*}{\textbf{TimerXL}}
& 0.3533 & 60.603 & 0.2704 & 34.140 & 0.2629 & 76.544 & 0.2554 & 44.796 & 0.3593 & 54.423 & 0.2700 & 65.115 & 0.3060 & 39.593 & {--} & {--}\\
& F & T  &
& 0.3215 & 56.152 & 0.3204 & 40.272 & 0.2766 & 74.769 & 0.1256 & 23.938 & 0.2629 & 43.267 & 0.1887 & 59.998 & 0.2799 & 34.378 & {--} & {--}\\
& T  & F &
& 0.2537 & 44.253 & 0.1582 & 20.996 & 0.1258 & 80.594 & 0.1834 & 32.811 & 0.2020 & 31.736 & 0.2713 & 61.718 & 0.2506 & 38.279 & {--} & {--}\\
& T  & T  &
& 0.2053 & 36.834 & 0.1302 & 18.081 & 0.1407 & 84.425 & 0.2005 & 31.594 & 0.1613 & 25.790 & 0.2730 & 63.362 & 0.2271 & 36.023 & {--} & {--}\\
\midrule

\multirow{3}{*}{\textbf{M2}} & & & \textbf{AutoTimes}
& 0.1154 & 21.596 & 0.1445 & 18.826 & 0.1258 & 63.496 & 0.1074 & 18.436 & 0.1091 & 19.734 & 0.1225 & 46.701 & 0.1519 & 23.847 & 0.1168 & 46.93\\
& & & \textbf{MAMBA}
& 0.1099 & 18.875 & \textbf{0.0958} & \textbf{13.370} & 0.1116 & 61.603 & 0.0510 & 8.9284 & \underline{0.1048} & \underline{18.240} & \underline{0.1001} & 43.372 & 0.1331 & 19.790 & 0.0721 & 44.78\\
& & & \textbf{TimerXL}
& 0.1635 & 29.288 & 0.1722 & 23.212 & 0.1720 & 90.708 & 0.1130 & 19.597 & 0.1550 & 27.510 & 0.1655 & 65.259 & 0.2066 & 31.550 & {--} & {--}\\
\midrule

\multirow{3}{*}{\textbf{Indep.}} & & & \textbf{AutoTimes}
& 0.1264 & 24.358 & 0.1583 & 23.841 & 0.1378 & 72.084 & 0.1177 & 20.095 & 0.1195 & 21.872 & 0.1342 & 48.363 & 0.1664 & 27.076 & {--} & {--}\\
& & & \textbf{TTM}
& 0.2408 & 41.839 & 0.1888 & 29.356 & 0.1662 & 41.195 & 0.1766 & 32.871 & 0.2087 & 30.259 & 0.1758 & \underline{24.414} & 0.2368 & 30.002 & {--} & {--}\\
& & & \textbf{MAMBA}
& \underline{0.0990} & \underline{18.845} & 0.1506 & 18.835 & 0.1196 & 52.032 & 0.0559 & 9.1538 & 0.1095 & 19.348 & 0.1092 & 28.615 & \underline{0.1199} & 16.821 & {--} & {--}\\
\midrule

\multicolumn{4}{l}{\textbf{NOCE-Net (Proposed)}}
& \textbf{0.0897} & \textbf{15.157}
& 0.1422 & 17.991
& \textbf{0.0574} & \textbf{14.606}
& \textbf{0.0415} & \textbf{7.7747}
& \textbf{0.0778} & \textbf{13.311}
& \textbf{0.0656} & \textbf{13.710}
& \textbf{0.0914} & \textbf{10.383}
& \textbf{0.0414} & \textbf{8.2325}\\
\bottomrule
\end{tabular}}
\vspace{0.5em}
\raggedright\scriptsize
\textbf{VS} = VARSEQ, \textbf{MP} = MPATCH, \textbf{F} = False, \textbf{T} = True, \textbf{M1} = Method 1, \textbf{M2} = Method 2, \textbf{Indep.} = Independent
\end{table*}

Table~\ref{tab:zero-shot} reports zero-shot and few-shot performance under three baseline configurations (averaged over 5 runs; std.\ $5$-$10\%$). For zero-shot evaluation we forecast the lifetime voltage response of the test batteries. \texttt{NOCE-Net} achieves $13.7\%$ MAPE and MAE of $0.0656$ in aggregate zero-shot evaluation—improvements of $43.8\%$ and $34.5\%$ over the next-best baselines (TTM + Indep. and MAMBA + M2 respectively)—while being $\sim 4\times$ smaller than the best and $40\times$ smaller than the average baseline (see Tab. \ref{tab:param_comp}). Transformer-based models - AutoTimes and TimerXL, generalise poorly to irregular sequences; MAMBA and TTM configurations with multi-patch embeddings handle variable-length inputs best among baselines but remain inferior overall. 

\begin{figure}[t]
    \centering
    \begin{minipage}[t]{0.49\textwidth}
        \centering
        \includegraphics[width=0.9\textwidth]{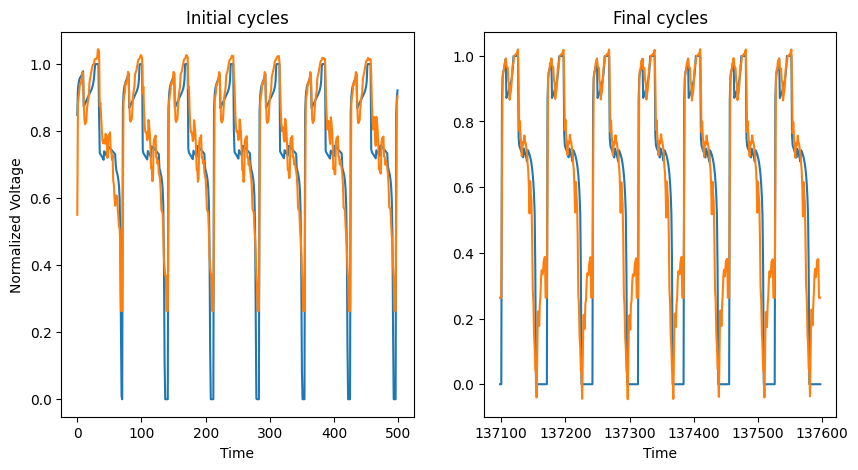}
        \subcaption{HUST}
        \label{fig:hust_pred}
    \end{minipage}
    \hfill
    \begin{minipage}[t]{0.49\textwidth}
        \centering
        \includegraphics[width=0.9\textwidth]{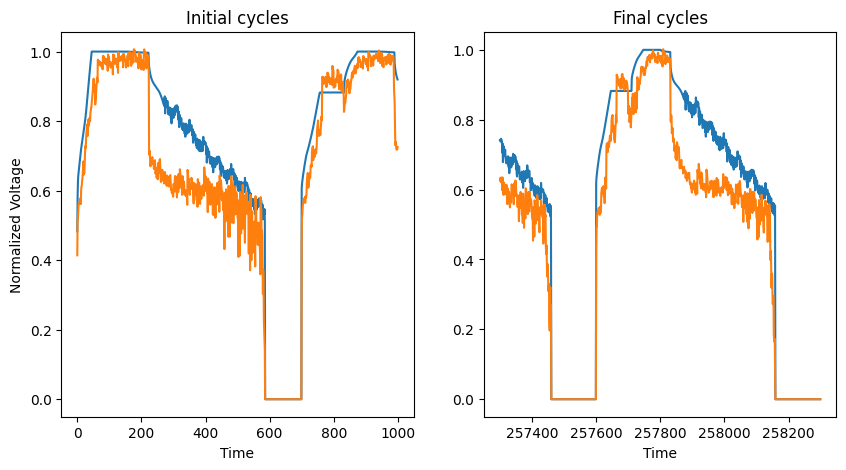}
        \subcaption{Stanford-EV}
        \label{fig:stanford_pred}
    \end{minipage}
    \caption{Predicted (Orange) vs True (Blue) voltage responses for randomly chosen batteries from the HUST and Stanford-EV datasets. For each dataset, the early life and towards end of life cycles are shown in the plots.}
    \label{fig:op_profs}
\end{figure}

For few-shot forecasting, we fine-tune only the final MLP heads of each model (ORL MLP for \texttt{NOCE-Net}) on the first 50 FECs per test cell; \texttt{NOCE-Net} retains the best aggregate performance, improving MAE by $42.5\%$ over MAMBA + M2. Figure~\ref{fig:op_profs} shows predicted vs.\ true voltage responses on two test datasets. Refer Tables \ref{tab:in-sample zero-shot} and \ref{tab:few-shot} (Appendix \ref{app:forecastres}) for complete per dataset results on in-sample datasets (comprising Total-Seen) and few-shot results on test datasets respectively.

\begin{table*}[t]
\caption{%
  Ablation study on \texttt{NOCE-Net} architectural components.
  Each value is the mean of 3 independent runs, std.\ shown in parentheses.
  Percentages give the relative change vs.\ full \texttt{NOCE-Net}
  (\textcolor{red}{$\uparrow$ red} = worse, \textcolor{green!55!black}{$\downarrow$ green} = better). Lower is better.
}
\label{tab:ablation}
\centering
\setlength{\tabcolsep}{3.5pt}
\renewcommand{\arraystretch}{1.0}
\resizebox{\textwidth}{!}{%
\small
\begin{tabular}{l cc cc cc cc cc cc}
\toprule
 & \multicolumn{2}{c}{\textbf{NASA-PCoE}}
 & \multicolumn{2}{c}{\textbf{Stanford-EV}}
 & \multicolumn{2}{c}{\textbf{HUST}}
 & \multicolumn{2}{c}{\textbf{XJTU-Full}}
 & \multicolumn{2}{c}{\textbf{evTOL}}
 & \multicolumn{2}{c}{\textbf{Total}} \\
\cmidrule(lr){2-3}\cmidrule(lr){4-5}\cmidrule(lr){6-7}
\cmidrule(lr){8-9}\cmidrule(lr){10-11}\cmidrule(lr){12-13}
 & MAE & MAPE
 & MAE & MAPE
 & MAE & MAPE
 & MAE & MAPE
 & MAE & MAPE
 & MAE & MAPE \\
\midrule

\textbf{NOCE-Net (full)}
 & 0.0897 & 15.16
 & 0.1422 & 17.99
 & 0.0574 & 14.61
 & 0.0415 & 7.77
 & 0.0778 & 13.31
 & 0.0656 & 13.71 \\[4pt]

\midrule

\textbf{wo Task Emb.}
 & \makecell{0.178 \\ \scriptsize{($\pm$0.010)} \\ \worse{98\%}}
 & \makecell{33.68 \\ \scriptsize{($\pm$1.85)} \\ \worse{122\%}}
 & \makecell{0.120 \\ \scriptsize{($\pm$0.006)} \\ \better{16\%}}
 & \makecell{16.97 \\ \scriptsize{($\pm$0.81)} \\ \better{6\%}}
 & \makecell{0.130 \\ \scriptsize{($\pm$0.005)} \\ \worse{127\%}}
 & \makecell{43.99 \\ \scriptsize{($\pm$1.54)} \\ \worse{201\%}}
 & \makecell{0.180 \\ \scriptsize{($\pm$0.008)} \\ \worse{334\%}}
 & \makecell{34.92 \\ \scriptsize{($\pm$1.47)} \\ \worse{349\%}}
 & \makecell{0.141 \\ \scriptsize{($\pm$0.007)} \\ \worse{81\%}}
 & \makecell{23.61 \\ \scriptsize{($\pm$1.09)} \\ \worse{77\%}}
 & \makecell{0.139 \\ \scriptsize{($\pm$0.005)} \\ \worse{112\%}}
 & \makecell{26.85 \\ \scriptsize{($\pm$1.02)} \\ \worse{96\%}} \\[10pt]

\textbf{wo FEC Emb.}
 & \makecell{0.179 \\ \scriptsize{($\pm$0.009)} \\ \worse{100\%}}
 & \makecell{33.60 \\ \scriptsize{($\pm$1.78)} \\ \worse{122\%}}
 & \makecell{0.120 \\ \scriptsize{($\pm$0.006)} \\ \better{16\%}}
 & \makecell{16.82 \\ \scriptsize{($\pm$0.84)} \\ \better{7\%}}
 & \makecell{0.130 \\ \scriptsize{($\pm$0.005)} \\ \worse{127\%}}
 & \makecell{43.84 \\ \scriptsize{($\pm$1.58)} \\ \worse{200\%}}
 & \makecell{0.185 \\ \scriptsize{($\pm$0.008)} \\ \worse{346\%}}
 & \makecell{34.57 \\ \scriptsize{($\pm$1.56)} \\ \worse{345\%}}
 & \makecell{0.142 \\ \scriptsize{($\pm$0.006)} \\ \worse{83\%}}
 & \makecell{23.58 \\ \scriptsize{($\pm$1.04)} \\ \worse{77\%}}
 & \makecell{0.140 \\ \scriptsize{($\pm$0.005)} \\ \worse{113\%}}
 & \makecell{26.72 \\ \scriptsize{($\pm$0.99)} \\ \worse{95\%}} \\[10pt]

\textbf{wo Learn Emb.}
 & \makecell{0.180 \\ \scriptsize{($\pm$0.010)} \\ \worse{101\%}}
 & \makecell{33.56 \\ \scriptsize{($\pm$1.95)} \\ \worse{121\%}}
 & \makecell{0.164 \\ \scriptsize{($\pm$0.008)} \\ \worse{15\%}}
 & \makecell{21.82 \\ \scriptsize{($\pm$1.00)} \\ \worse{21\%}}
 & \makecell{0.130 \\ \scriptsize{($\pm$0.005)} \\ \worse{127\%}}
 & \makecell{43.80 \\ \scriptsize{($\pm$1.66)} \\ \worse{200\%}}
 & \makecell{0.185 \\ \scriptsize{($\pm$0.008)} \\ \worse{346\%}}
 & \makecell{34.42 \\ \scriptsize{($\pm$1.48)} \\ \worse{343\%}}
 & \makecell{0.142 \\ \scriptsize{($\pm$0.007)} \\ \worse{83\%}}
 & \makecell{23.57 \\ \scriptsize{($\pm$1.11)} \\ \worse{77\%}}
 & \makecell{0.143 \\ \scriptsize{($\pm$0.006)} \\ \worse{118\%}}
 & \makecell{27.13 \\ \scriptsize{($\pm$1.09)} \\ \worse{98\%}} \\[10pt]

\textbf{wo Rec. Backbone}
 & \makecell{0.167 \\ \scriptsize{($\pm$0.011)} \\ \worse{86\%}}
 & \makecell{30.64 \\ \scriptsize{($\pm$1.99)} \\ \worse{102\%}}
 & \makecell{0.140 \\ \scriptsize{($\pm$0.008)} \\ \better{2\%}}
 & \makecell{18.19 \\ \scriptsize{($\pm$1.06)} \\ \worse{1\%}}
 & \makecell{0.154 \\ \scriptsize{($\pm$0.007)} \\ \worse{168\%}}
 & \makecell{44.30 \\ \scriptsize{($\pm$1.99)} \\ \worse{203\%}}
 & \makecell{0.175 \\ \scriptsize{($\pm$0.010)} \\ \worse{322\%}}
 & \makecell{31.65 \\ \scriptsize{($\pm$1.74)} \\ \worse{307\%}}
 & \makecell{0.134 \\ \scriptsize{($\pm$0.007)} \\ \worse{72\%}}
 & \makecell{22.32 \\ \scriptsize{($\pm$1.16)} \\ \worse{68\%}}
 & \makecell{0.160 \\ \scriptsize{($\pm$0.008)} \\ \worse{144\%}}
 & \makecell{36.84 \\ \scriptsize{($\pm$1.77)} \\ \worse{169\%}} \\

\bottomrule
\end{tabular}}
\end{table*}


\subsection{Ablation Studies}\label{sec:res_ablation}

\begin{figure}
  \vspace{-6pt}
  \centering
  \begin{minipage}[t]{0.45\textwidth}
    \centering
    \includegraphics[width=\textwidth]{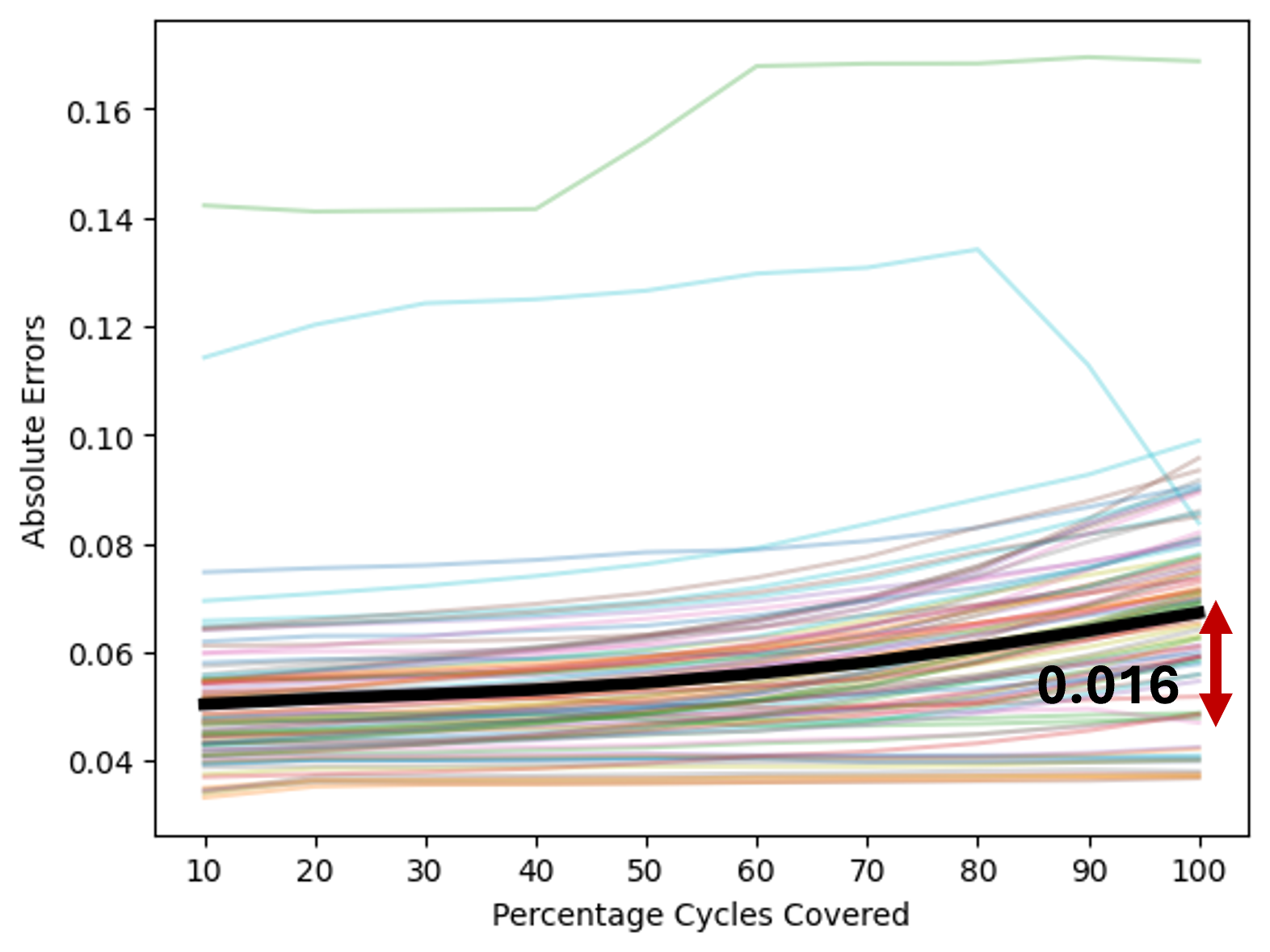}
    \subcaption{Without recurrence}
    \label{fig:wo_rec}
  \end{minipage}
  \hfill
  \begin{minipage}[t]{0.45\textwidth}
    \centering
    \includegraphics[width=\textwidth]{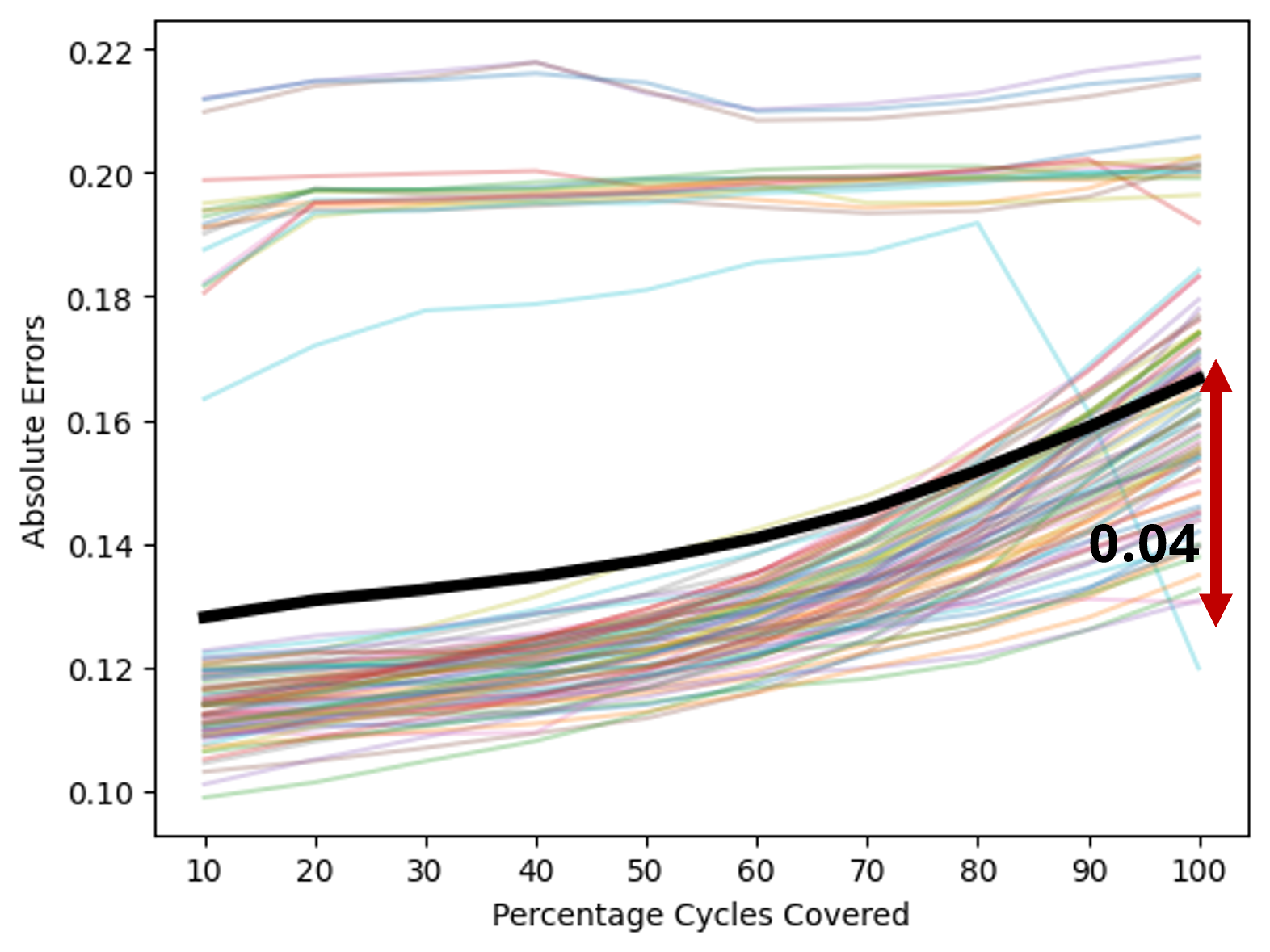}
    \subcaption{Full model}
    \label{fig:wrec}
  \end{minipage}
  \caption{MAE across FEC\%. Without recurrence, error grows $3\times$ faster with cycle index.}
  \label{fig:wo_and_with_rec}
  \vspace{-8pt}
\end{figure}
Table~\ref{tab:ablation} ablates core components of \texttt{NOCE-Net}. Removing task, FEC, and learnable-token embeddings degrades aggregate zero-shot MAE by $\sim50\%$, confirming their role in conditioning the model on heterogeneous chemistries and operating conditions. Eliminating the recurrent backbone causes a substantially larger collapse (
$MAE=0.16$, $MAPE=36.84\%$
), establishing that inter-cycle dependency modelling is essential for long-horizon voltage prediction. Figure~\ref{fig:wo_and_with_rec} plots the MAE across cycle \% till EOL for each of the test batteries with and without recurrence. It shows that without recurrence, intra-cycle prediction error accumulates with FEC index at nearly $3\times$ the rate of the full model, demonstrating that recurrent inter-cycle representations are necessary to track how aging-driven shifts in battery dynamics modulate the voltage response over the full operational lifetime.

\subsection{Transferability of Learned Representations to SOH Prediction}
\label{sec:res_soh_transferability}

NOCE-Net is trained exclusively on voltage response prediction, yet the GRU hidden states, $\mathcal{H} = \{\boldsymbol h_c\}_{c\in [N]}$, accumulate operational context over the battery lifetime
and may encode degradation-relevant information transferable to downstream tasks such as state-of-health~(SOH) estimation. We represent SOH as normalized capacity fade, i.e., $SOH = (Q - 0.8Q_{nom})/(Q_{nom} - 0.8Q_{nom})$, where $Q_{nom}$ is the nominal capacity and $0.8Q_{nom}$ is the largely accepted EOL capacity limit.
(Full numerical results and experimental details are provided in Appendix~\ref{app:soh_analysis}.)

\begin{figure*}[t]
  \centering
  \includegraphics[width=\textwidth]{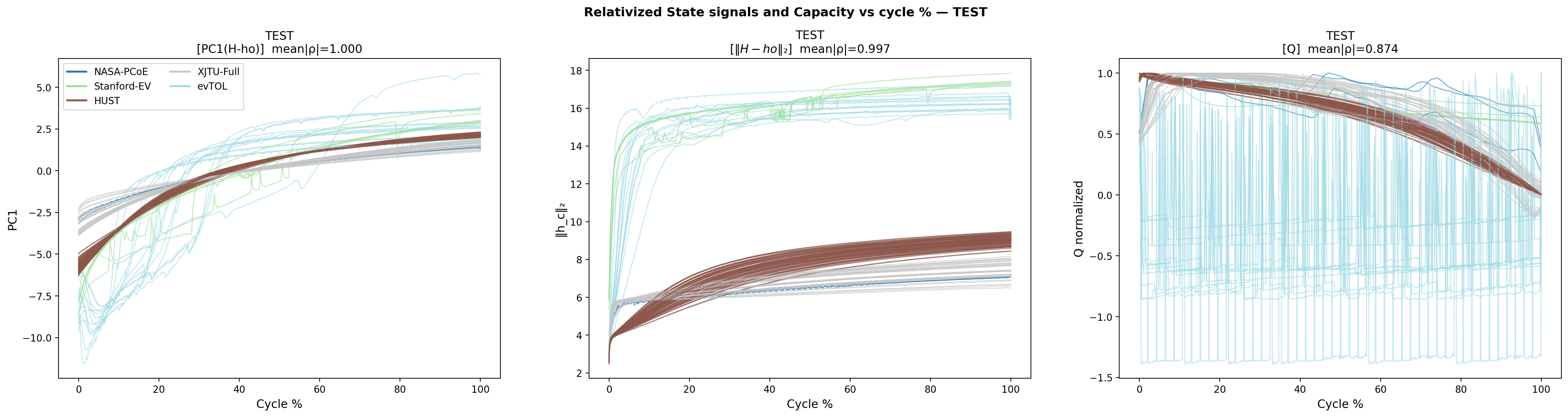}\\[3pt]
  \includegraphics[width=\textwidth]{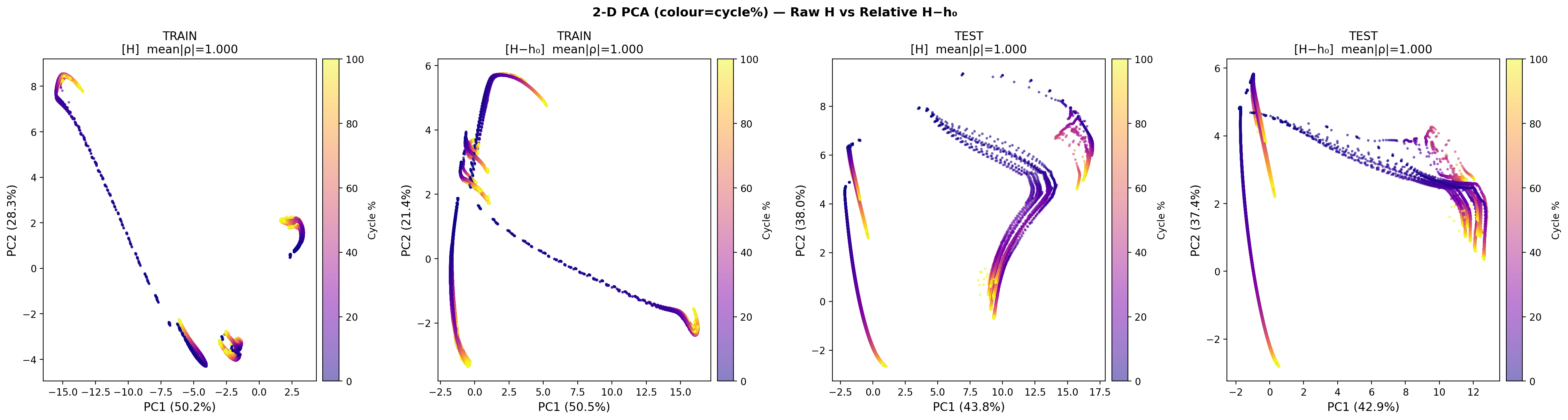}
  \caption{%
    GRU hidden state representation diagnostics.
    \textit{Top}: PC1, L2 norm, and normalised SOH vs.\ cycle\% per test battery. 
    All summaries of $H{-}h_0$ evolve monotonically
    (mean $|\rho| \sim 1$).
    \textit{Bottom}: 2-D PCA trajectory per battery; filled dots mark BOL.
    Under raw $H$ arcs have inconsistent orientations; under $H{-}h_0$
    all arcs share consistent degradation directions
    across all chemistries.
  }
  \label{fig:representation_diagnostics}
\end{figure*}

\textbf{Rank Correlation: Hidden states encode degradation indicating signals}
For each battery in $\texttt{TRAIN}\cup\texttt{TEST}$, we compute the Spearman $|\rho|$
between the FEC index $c$ and four scalar summaries of the hidden state sequence - the L2 norm~$\|\boldsymbol h_c\|_2$, the first principal component~$\mathrm{PC1}(\mathcal{H})$,
and their relativised counterparts, $\|\boldsymbol h_c-\boldsymbol h_0\|_2$ and $\mathrm{PC1}(\mathcal{H}-\boldsymbol h_0)$, obtained by subtracting the
battery-specific initialization $\boldsymbol h_0$ (Ref. Table \ref{tab:monotonicity_diagnostics}, Appendix \ref{app:spearman}).
$\mathrm{PC1}(\mathcal{H})$ achieves mean $|\rho| = 1.000$
against the FEC index and $|\rho| \geq 0.83$ against normalized capacity fade (SOH) across all datasets with monotone degradation (except for evTOL and NASA-PCoE which have erratic capacity fade signals; see Fig.~\ref{fig:representation_diagnostics} and \ref{fig:full_diagnostics}).
The raw norm, $\|\boldsymbol h_c\|_2$, achieves only $|\rho| = 0.750$ on zero-shot test batteries, recovering to $0.997$ after relativisation — revealing that the task-conditioned initialisation $\boldsymbol h_0$ introduces a battery-specific offset orthogonal to the degradation direction, which is eliminated by $\boldsymbol h_c - \boldsymbol h_0$ without any additional supervision.
Figure~\ref{fig:representation_diagnostics} illustrates this geometrically: relativised states trace monotone arcs in PCA space for both training and zero-shot batteries, while raw states show no consistent direction. All scalar summaries are monotone against cycle $\%$. evTOL and NASA-PCoE have erratic SOH curves. Intriguingly, the PCA arcs bear visual resemblance to capacity fade curves with characteristic knee regions — an observation we leave for future investigation.


\textbf{Shape Decomposition: Latents capture degradation shape,
remaining errors are affine.}
To understand the relation between SOH and the recurrent states, we train a lightweight MLP for SOH prediction jointly on the training dataset's frozen hidden states (ref. complete in-sample and test dataset results in Table \ref{tab:direct_soh}). We extract Proportional, Integral, Difference and Residual(PID-R) features (Appendix~\ref{app:pid_features}) from the hidden states for this purpose. On both in sample and zero shot test batteries, we compute Pearson shape correlation~$r$ between the centred predicted and true
SOH trajectories, alongside the signed mean offset
$\bar{\delta} = \overline{\hat{SOH}} - \bar{SOH}$ and the scale ratio
$\sigma_{SOH} / \sigma_{\hat{SOH}}$, and average them per battery dataset. We observe consistently high shape correlations (mean $r = 0.856$, range $0.69$--$0.99$, evTOL excluded), confirming that the recurrent state latents encode capacity fade \emph{shape} without any capacity supervision.
The dominant error is affine, with signed offsets spanning $[-0.78,\, +0.65]$~SOH units and scale
ratios spanning $[0.37,\, 5.23]$, varying \emph{within} datasets (std of scale ratio within MATR: $0.21$; within
ISU\_ILCC: $0.42$) rather than being fixed per chemistry — indicating intra-dataset fade-rate variability rather than systematic model bias. (See appendix \ref{app:soh_head} for details.)

\textbf{Ruling out cycle progression as a confound.}
A natural concern is that the correlation between the hidden-state
representations and capacity could simply reflect cycle progression
rather than a genuinely learned degradation signal. We test this with FEC-only baselines — a linear model and a 3-layer MLP mapping only the normalised FEC index (no hidden state) to SOH — against the PID-R $\to$SOH-head pipeline (Table~\ref{tab:fec_baseline}). On zero-shot batteries, PID-R achieves $52.7\%$ and $90.5\%$ lower aggregate MAPE than the FEC-only Linear and MLP baselines respectively; a nonlinear
model with direct access to the FEC index still cannot match the
frozen hidden-state representation. The FEC-only baselines only edge
out PID-R on batteries where the SOH head itself exhibits known affine
miscalibration (ISU\_ILCC, NASA~Rand, Stanford-EV; Table~\ref{tab:shape_diagnostic}),
not because cycle progression is inherently more predictive, confirming the hidden states encode battery-specific degradation structure beyond
simple throughput tracking.

\textbf{Per-dataset fine-tuning substantially reduces errors.}
We evaluate how much of the residual affine error can be corrected by
training a per-dataset MLP head on a fraction of in-domain batteries,
using the frozen PID-R features as input (Appendix~\ref{app:per_dataset_mlp}).
Table~\ref{tab:mlp_30pct} summarises results at 30\% training fraction per battery. We repeat this for 10, 20, 40 and 50 \% training fractions as well and Table~\ref{tab:mlp_full_sweep} contains the full sweep.
MATR, Stanford-EV, HUST, and XJTU-Full achieve
$5$--$33\%$ MAPE at 30\%, representing substantial improvements over the joint SOH training baseline (compare with Table \ref{tab:direct_soh}).
ISU\_ILCC and NASA~Rand remain the hardest cases at $72\%$ and $77\%$ respectively, attributable to very slow fade rates (scale needed $\approx 2.5$ and $1.3$), high intra-dataset variability from
variable partial-cycling protocols, and small absolute SOH changes per cycle that amplify MAPE for deeply degraded batteries.


\begin{table*}[t]
\caption{%
  Per-dataset MLP head SOH prediction at 30\% training fraction
  (3 random splits; evTOL and UMich excluded).
  Full sweep in Table~\ref{tab:mlp_full_sweep}.
}
\label{tab:mlp_30pct}
\vspace{2pt}
\centering
\setlength{\tabcolsep}{4pt}
\renewcommand{\arraystretch}{0.95}
\resizebox{\textwidth}{!}{%
\small
\begin{tabular}{l
  cc cc cc cc cc cc cc cc}
\toprule
& \multicolumn{2}{c}{\textbf{Stanford-EV}}
& \multicolumn{2}{c}{\textbf{MATR}}
& \multicolumn{2}{c}{\textbf{HUST}}
& \multicolumn{2}{c}{\textbf{XJTU-Full}}
& \multicolumn{2}{c}{\textbf{XJTU-Partial}}
& \multicolumn{2}{c}{\textbf{TJU}}
& \multicolumn{2}{c}{\textbf{ISU-ILCC}}
& \multicolumn{2}{c}{\textbf{NASA Rand}} \\
\cmidrule(lr){2-3}\cmidrule(lr){4-5}\cmidrule(lr){6-7}\cmidrule(lr){8-9}
\cmidrule(lr){10-11}\cmidrule(lr){12-13}\cmidrule(lr){14-15}\cmidrule(lr){16-17}
& MAPE & MAE
& MAPE & MAE
& MAPE & MAE
& MAPE & MAE
& MAPE & MAE
& MAPE & MAE
& MAPE & MAE
& MAPE & MAE \\
\midrule
& 5.5\% & 0.043
& 21.6\% & 0.076
& 30.9\% & 0.074
& 32.1\% & 0.070
& 39.0\% & 0.158
& 51.3\% & 0.128
& 71.4\% & 0.355
& 76.6\% & 0.637 \\
& \scriptsize{$\pm$0.6\%} & \scriptsize{$\pm$0.004}
& \scriptsize{$\pm$1.7\%} & \scriptsize{$\pm$0.006}
& \scriptsize{$\pm$2.6\%} & \scriptsize{$\pm$0.007}
& \scriptsize{$\pm$1.9\%} & \scriptsize{$\pm$0.005}
& \scriptsize{$\pm$9.5\%} & \scriptsize{$\pm$0.031}
& \scriptsize{$\pm$0.7\%} & \scriptsize{$\pm$0.003}
& \scriptsize{$\pm$1.7\%} & \scriptsize{$\pm$0.014}
& \scriptsize{$\pm$13.2\%} & \scriptsize{$\pm$0.063} \\
\bottomrule
\end{tabular}}
\end{table*}


\section{Conclusion and Limitations}\label{sec:conclusion}

This work introduced \texttt{NOCE-Net}, a control-enabling, non-invasive framework for long-horizon battery voltage forecasting that couples intra-cycle operational inputs with inter-cycle latent dynamics to capture degradation-driven changes in battery behaviour without RPTs. Extensive zero-shot and few-shot evaluations across diverse chemistries demonstrate consistent outperformance of strong baselines at substantially lower parameter cost. Ablation studies confirm that task and FEC conditioning and the recurrent backbone are each critical for learning the long-term dynamics that govern voltage response over battery life. Beyond voltage forecasting, the recurrent hidden states encode battery degradation structure without capacity supervision, transferring to SOH estimation with near-affine residual error and strong gains from light per-dataset calibration (Section \ref{sec:res_soh_transferability}).

However, several limitations bound the current work. Voltage supervision encodes degradation \emph{direction, shape} but not absolute \emph{scale}, requiring per-dataset calibration - joint training with an SOH auxiliary loss is a natural remedy. evTOL-style partial-discharge profiles produce non-monotone capacity trajectories for which scalar SOH is ill-defined. Moreover, re-BatteryData requires current, temperature, and time signals: datasets lacking temperature records will require imputation. Finally, the sequential GRU backbone is a computational bottleneck relative to fully parallel architectures. Future extensions of this work include developing solutions addressing the open challenges as well as probing \texttt{NOCE-Net} as a surrogate for data-driven battery optimisation in vehicular and grid applications, the closed-loop deployment of which requires controllable or reliably forecasted temperature. Stronger real-world validation under such settings therefore remains an open challenge.
\newpage
\appendix

\section{Related Work}\label{app:relatedwork}

\subsection{Feature-Based Data-Driven Battery Modeling}\label{appsec:featuremodels}
A large body of work in data-driven battery modeling relies on manually engineered features extracted from battery operational inputs and measured response signals, particularly voltage–capacity ($Q$–$V$) curves and their derivatives. Seminal work by .~\cite{Severson2019} demonstrated that statistical features derived from early-cycle differential voltage ($dV/dQ$) curves can predict battery lifetime well before observable capacity fade. This paradigm has since been extended through the use of partial charging curves, fragmented charge segments, fixed voltage windows, and hand-crafted statistics for tasks such as state-of-health (SOH) estimation and remaining useful life (RUL) prediction~\cite{Lu2023, Zhu2022, Richardson2019, Zhang12025}.

While effective in controlled laboratory settings, these approaches are fundamentally response-dependent: feature construction requires access to voltage trajectories, which are unavailable for unseen future operating conditions. Consequently, such models cannot be evaluated a priori under hypothetical input trajectories and are therefore unsuitable for predictive optimization or control-oriented tasks. To improve robustness, many studies deliberately restrict feature extraction to short, repeatable operational segments (e.g., constant-current charging phases or partial charging curves assumed to be consistently present (\cite{Zhang12025, Zhu2022})). This necessitates additional event-detection heuristics and discards large portions of the operational history, leaving long-term usage patterns and path-dependent effects unmodeled.

Large-scale efforts such as BatteryLife~\cite{10.1145/3711896.3737372} significantly expand dataset diversity across chemistries and form factors, yet retain an early-life prediction formulation: end-of-life is inferred using the first $\sim$100 cycles of input–output data. Such formulations implicitly assume that operating conditions remain stationary over the battery lifetime and do not support control-enabling modeling, where future behavior must be predicted under evolving operational regimes. As a result, despite increased dataset scale, these approaches remain limited to retrospective or static prediction settings.

\subsection{Hybrid Physics–Data-Driven Battery Models}\label{appsec:hybridmodels}

Hybrid modeling approaches combine analytical battery models with machine learning components to balance interpretability and flexibility. Physics-informed neural networks (PINNs) embed neural modules within equivalent circuit models (ECMs) or electrochemical models to learn time-varying parameters and degradation dynamics~\cite{10310297, Wang2024, Nascimento2023-or}. Other works employ Gaussian process–based circuit models~\cite{ZHOU2025126375}, Bayesian formulations~\cite{Jones2022-lf}, or neural parameterizations of electrochemical models to improve voltage or capacity prediction accuracy.

These methods offer strong physical grounding and interpretability, but their empirical validation is typically limited to small numbers of batteries and narrow operational regimes, often within a single dataset such as NASA or HUST. Moreover, many hybrid approaches rely on recurrent or ODE-based latent dynamics, introducing significant computational overhead that limits scalability to long, high-resolution operational histories. In addition, most hybrid models require periodic Reference Performance Tests (RPTs) to measure capacity fade or impedance growth, which necessitate taking batteries offline and disrupt normal operation~\cite{Lanubile2024, POZZATO20232035}. These constraints make hybrid methods ill-suited for universal, lifetime-scale representation learning across heterogeneous datasets and operational contexts.

\subsection{Generalization, Transfer Learning, and Domain Adaptation in Battery Modeling}\label{appsec:tlmodels}

To address dataset heterogeneity and limited data availability, several studies explore transfer learning\cite{Ma2022, Zhang2025}, continual learning \cite{Che2023-dl}, and unsupervised domain adaptation \cite{Lu2023}for battery health estimation. These methods typically adapt models trained on source batteries to new target batteries by fine-tuning on small amounts of labeled data or aligning feature distributions across domains.

However, demonstrated generalization gains are often confined to small transfer gaps, such as batteries of similar chemistries, form factors, and cycling protocols. Furthermore, most such methods still rely on hand-crafted response-dependent features (e.g., $Q$–$V$ or $dV/dQ$ curves), inheriting their fundamental limitations for control-enabling tasks. As a result, while transfer learning improves robustness within narrow regimes, it does not address the challenge of learning universal representations that remain valid under evolving operational conditions and across diverse datasets\cite{LIU2023100117}.

\subsection{Large-Scale Battery Datasets and Data Standardization}\label{appsec:largedatasets}

Recent efforts have begun addressing data scarcity by aggregating large collections of public battery datasets. BatteryML~\cite{zhang2024batteryml} and BatteryLife~\cite{10.1145/3711896.3737372} propose standardized data formats and benchmarks for SOH and lifetime prediction. These initiatives represent important steps toward reproducibility and scale.

Nevertheless, their data pipelines are primarily designed for retrospective analysis and static prediction tasks. The BatteryData structure proposed in ~\cite{zhang2024batteryml} is used primarily for effective feature extraction and SOH or RUL prediction using off the shelf ML models. ~\cite{zhang2024batteryml} collects a much larger dataset of 900 batteries and develops a $CyclePatch$ layer that facilitates RUL prediction using early life data. It uses a fixed-length resampling mechanism on early life data to convert it into a uniform temporal dimension data. Moreover, both these works typically assume semantically non-identical cycle definitions, fixed or uniform sampling after preprocessing, and flat sequence structures, which limit their suitability for end-to-end representation learning from raw operational context.

\subsection{Foundational Time-Series Models and Structural Limitations}\label{appsec:foundationmodels}

Inspired by advances in language and vision, recent work has proposed foundation models\cite{shi2025timemoe} for time series forecasting\cite{liu2025timerxl}, anomaly detection\cite{wu2023timesnet}, and classification\cite{NEURIPS2024_0113ef46}. \cite{Wen2023} summarize transformer-based architectures and their adaptations for time-series data. These models typically assume uniformly sampled data, fixed-length contexts, or a single continuous temporal dimension. To handle irregularity, such models often rely on zero-padding, truncation, interpolation, or flattening of hierarchical structure into a single sequence. When applied to battery operational data—where each outer time step (cycle) may itself be a time series of variable length—these assumptions lead to significant information loss. Classical recurrent models (RNNs, LSTMs, GRUs)\cite{chung2014empiricalevaluationgatedrecurrent, 6795963} naturally support variable-length inputs but do not scale well to extremely long sequences and lack explicit mechanisms for hierarchical or nested sequence modeling. Consequently, existing foundational time-series models are not architecturally optimized for nested irregular sequence structures.

\section{Methodology}\label{app:methodology}
\subsection{\texttt{re-BatteryData}}
\subsubsection{Uniform Feature Dimension}\label{app:uniform_feats}
The raw aging datasets need to have records for three fields – current ($\mathbf{I} \in \mathbb{R}^L$), temperature and time($\mathbf{t}$) – to compute all the six features using \texttt{re-BatteryData}. $\mathbf{I}_{chg} = \mathbf{I}[\mathbf{I}>0]$, $\mathbf{I}_{dch} = \mathbf{I}[\mathbf{I}<0]$. The charge and discharge throughput is computed using $ChargeAccumulator()$ that takes in the raw current signal $\mathbf{I} \in \mathbb{R}^L $ and the sampling interval vector, $\mathbf{\Delta t} = \mathbf{t}[1:] - \mathbf{t}[:-1] \in \mathbb{R}^{L-1} $ to determine the in-cycle (dis)charge throughput. Let $\mathcal{K} = [k_1, k_2, \ldots, k_S]$ represent the index set corresponding to successive sign changes in $\mathbf{I}$, then, the in-cycle (dis)charge throughput $\mathbf{Q}\in \mathbb{R}^L$ is formally given by:
    \begin{align*}
        & \mathbf{Q} = ChargeAccumulator(\mathbf{I}, \mathcal{K})\\
        & \mathbf{Q}[i] = \sum_{j=k*}^{i} |\mathbf{I}[j]| \mathbf{\Delta t}[j] \\
        & k* = min_{k \in \mathcal{K} ; k < i} |k-i|
    \end{align*}
    It computes the sum of the absolute value of the current multiplied by the time interval for successive charge and discharge operations. Whenever the sign of the current changes, the charge accumulation operation resets and starts integrating the current from scratch till the next change in the sign of the current. $\mathbf{Q}_{chg} = \mathbf{Q}[\mathbf{I}>0]$ and $\mathbf{Q}_{dch} = \mathbf{Q}[\mathbf{I}<0]$ are used as two different features. It is not necessary for the batteries to have accurate records of these features. For instance, a number of the test and training batteries used in this work recorded the ambient temperature, which is easier to measure, instead of the battery temperature. Similarly, the chemistry of the evTOL batteries are completely unknown.\\
\subsubsection{Battery Specific Task Features}\label{app:task_feats}

\begin{table*}[t]
\caption{Battery Specific Task Features used for task embeddings}
\label{tab1}
\resizebox{\textwidth}{!}{%
\small
\begin{tabular}{|l|c|c|}

\hline
\multicolumn{1}{|c|}{\textbf{Categorical  Task Feature}} & \textbf{Classes}                                  & \textbf{Number of   Classes} \\ \hline
Form Factor                                              & Pouch,   Prismatic, Cylindrical, Polymer, Unknown & 5                            \\ \hline
Cathode   Composition                                      & LFP, LCO, NCA,   NMC (NCM), NCA + NCM, Unknown    & 6                            \\ \hline
Anode   Composition                                    & Graphite,  Graphite/Silicon, Unknown              & 3                            \\ \hline
\end{tabular}}
\end{table*}

To model variability across battery chemistries and configurations, we also record battery specific task features derived from battery metadata, specifically- Nominal Capacity ($Q_{nom}$), Voltage Range ($V_{max}$, $V_{min}$), Form Factor (e.g., pouch, cylindrical, etc.) and Electrode Composition (anode/cathode materials). Form factor, Anode and cathode composition are modelled as categorical features with 5, 6 and 3 categories. Nominal capacity and manufacturer specified voltage limits are continuous features. Each categorical feature has an additional ‘Unknown’ class to account for any batteries with unspecified or unknown metadata as shown in Table \ref{tab1}.	\\
\subsubsection{Full Equivalent Cycle based Segmentation}\label{app:FEC_importance}
\begin{figure*}[t]
    \centering
        \centering
        \includegraphics[width=\textwidth]{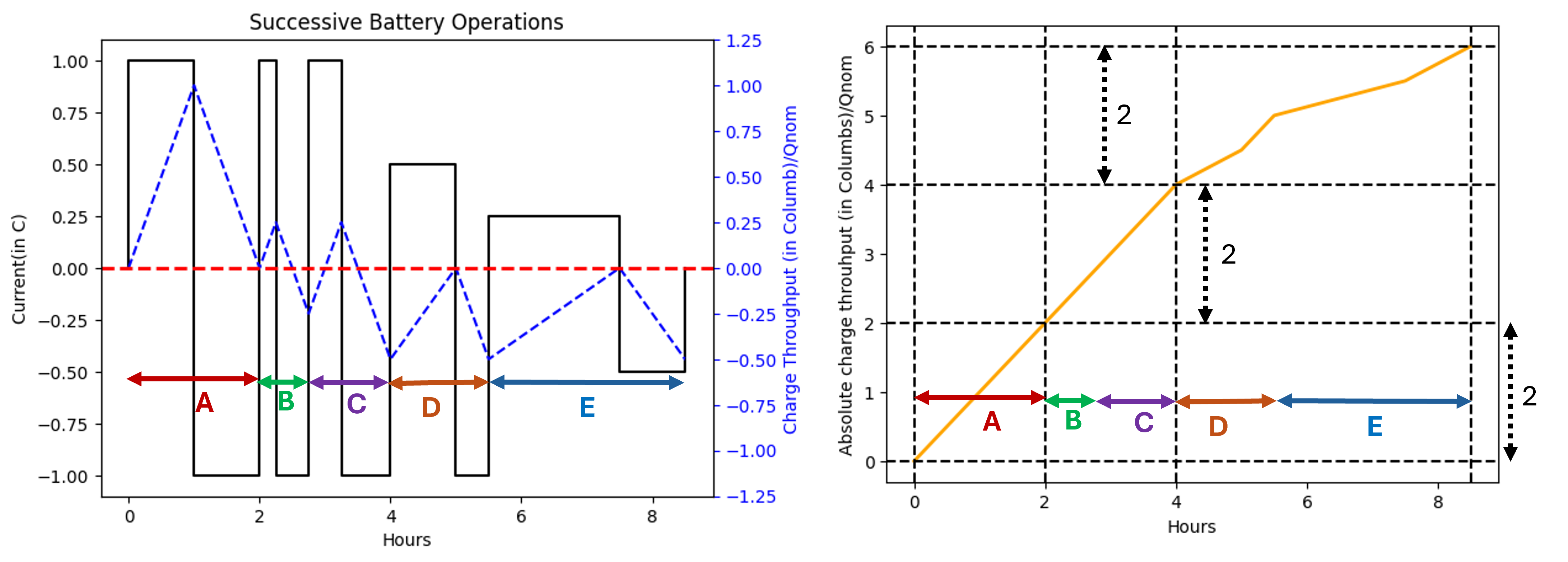}
        \caption{Five successive charge-discharge cycles are depicted. (a) represents the current profile for the cycles and the corresponding charge throughput and (b) represents the absolute charge throughput for the entire duration. Using simple counting based mechanisms might consider these cycles as separate equivalent entities while full cycling definitions don't account for cycles B-E. A semantically equivalent cycling definition is proposed through absolute charge throughput.}
        \label{fig:1}
\end{figure*}

A lot of interesting battery dynamics, like degradation, arise over a longer timescale compared to battery response. To model these dynamics, battery operational data often needs to be segregated into fundamental operational segments called cycles, which traditionally represent a period of successive charging and discharging operations. A popular way of defining an operational cycle is through a full cycle which  represents a period when the battery has undergone a full charge ($0-100\%$ SOC) and discharge ($100-0\%$ SOC) successively. Other ways include defining a cycle as a consecutively occuring charge discharge operation. However, any mechanism that simply counts the number of successive charge-discharge operations to determine an operational segment falls short of creating a semantically equivalent operation-agnostic scale for modelling such secondary slow moving dynamics like degradation. This is because the duration and number of charge-discharge cycles in a specific period of time directly depend on the specific battery operation, as explained in Fig.\ref{fig:1}, which depicts five successive charge-discharge operations denoted as A, B, C, D and E. 
A is a full charge-discharge cycle, while the others are partial cycles. Cycles B and C operate with a constant 1C current, while D and E operate with varying current levels. If we use a standard charge-discharge cycle counting mechanism as used in \cite{zhang2024batteryml, 10.1145/3711896.3737372}, then all of them will be considered equivalent cycles, whereas they are semantically distinct. Using the full-cycle definition to denote an operational cycle does not provide a way to model operations in cycles B-E.\\
To alleviate this problem, we propose using a semantically equivalent operation-agnostic operational cycle definition based on absolute charge throughput aggregation. Formally, for any two points $t_1$ and $t_2$ in time, we define a quantity $F_Q$ as:
\begin{align}
    F_Q = \frac{\int_{t_1}^{t_2}|I_t| dt}{Q_{nom}}
\end{align}
where $Q_{nom}$ is the nominal capacity of a battery and $I_t$ represents the current input to the battery. The quantity in the numerator is the absolute charge throughput or $Q_{acc} = \int_{t_1}^{t_2}|I_t| dt$. We quantify $[t_1, t_2]$ as a period of an operational cycle or a full equivalent cycle if $F_Q = 2$, i.e., the battery undergoes a total absolute charge throughput of twice its nominal capacity during that period. Accordingly, in Figure \ref{fig:1}, there are three operational cycles or full equivalent cycles (FECs), given by A, (B,C) and (D,E) because they represent the same normalized absolute charge throughputs. This definition is a generalized version of the full cycle definition. 

\section{Datasets}\label{app:datasets}
The training datasets comprise of:
\begin{enumerate}
    \item ISU-ILCC \cite{https://doi.org/10.25380/iastate.22582234}: Comprises of 155 polymer NCM batteries subjected to variable charge/discharge profiles.
    \item NASA-Rand \cite{nasaRandomizedBattery}: The NASA Random dataset comprises cylindrical NMC–Graphite cells subjected to randomized discharge profiles, providing partially cycled data across 13 cells.
    \item MATR \cite{Severson2019}: The MATR dataset, based on cylindrical LFP–Graphite cells, includes 124 full-cycle sequences under constant discharge and multi-step charge profiles, offering long-duration, high-resolution data.
    \item XJTU-Partial \cite{Wang2024}: The XJTU Partial dataset features NCM–Graphite cylindrical cells operated under controlled constant charge–discharge conditions with partial cycling across 16 cells.
    \item TJU \cite{Zhu2022} consists of 100 batteries with a mix of NCA, NCM and NCA+NCM batteries operated at different temperatures. 
    \item UMICH \cite{https://doi.org/10.7302/7tw1-kc35} consists of 21 NMC Pouch batteries tested under mechanical stress and conditions to observed degradation.
\end{enumerate}
   For out-of-distribution evaluation, five additional datasets were used:
   \begin{enumerate}
   \item eVTOL \cite{Bills2021}, representing aircraft-level variable discharge operations
   \item NASA PCoE \cite{SahaGoebel2007NASA}, a full-cycle NMC–Graphite dataset with moderate depth of discharge
   \item Stanford EV \cite{Pozzato2022}, capturing EV-oriented dynamic charging/discharging behavior in NMC and Graphite/Silicon cells.
   \item HUST \cite{Ma2022}, a large-scale LFP–Graphite dataset with extensive full-cycle data (77 cells).
   \item XJTU Full \cite{Wang2024}, another NCM–Graphite dataset representing full dynamic cycles over 39 cells.
   \end{enumerate}
   
\begin{figure*}[t]
    \centering
        \centering
        \includegraphics[width=\textwidth]{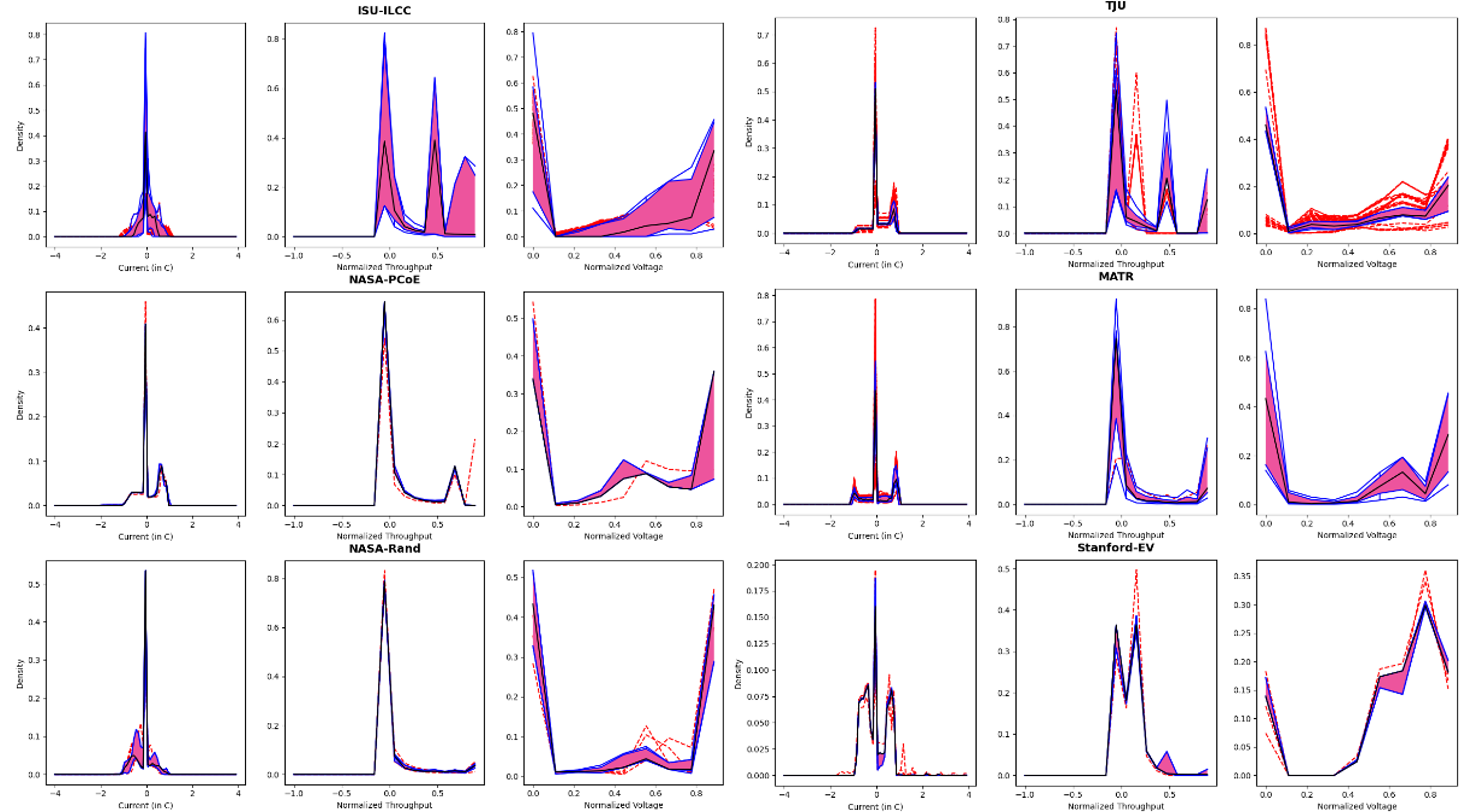}
        \caption{Functional box plots demonstrating the differences between the distributions of current, throughput and volatge in each of the battery datasets used in this work.}
        \label{fig:fb_plots}
\end{figure*}

\begin{figure*}[t]
    \centering
        \centering
        \includegraphics[width=\textwidth]{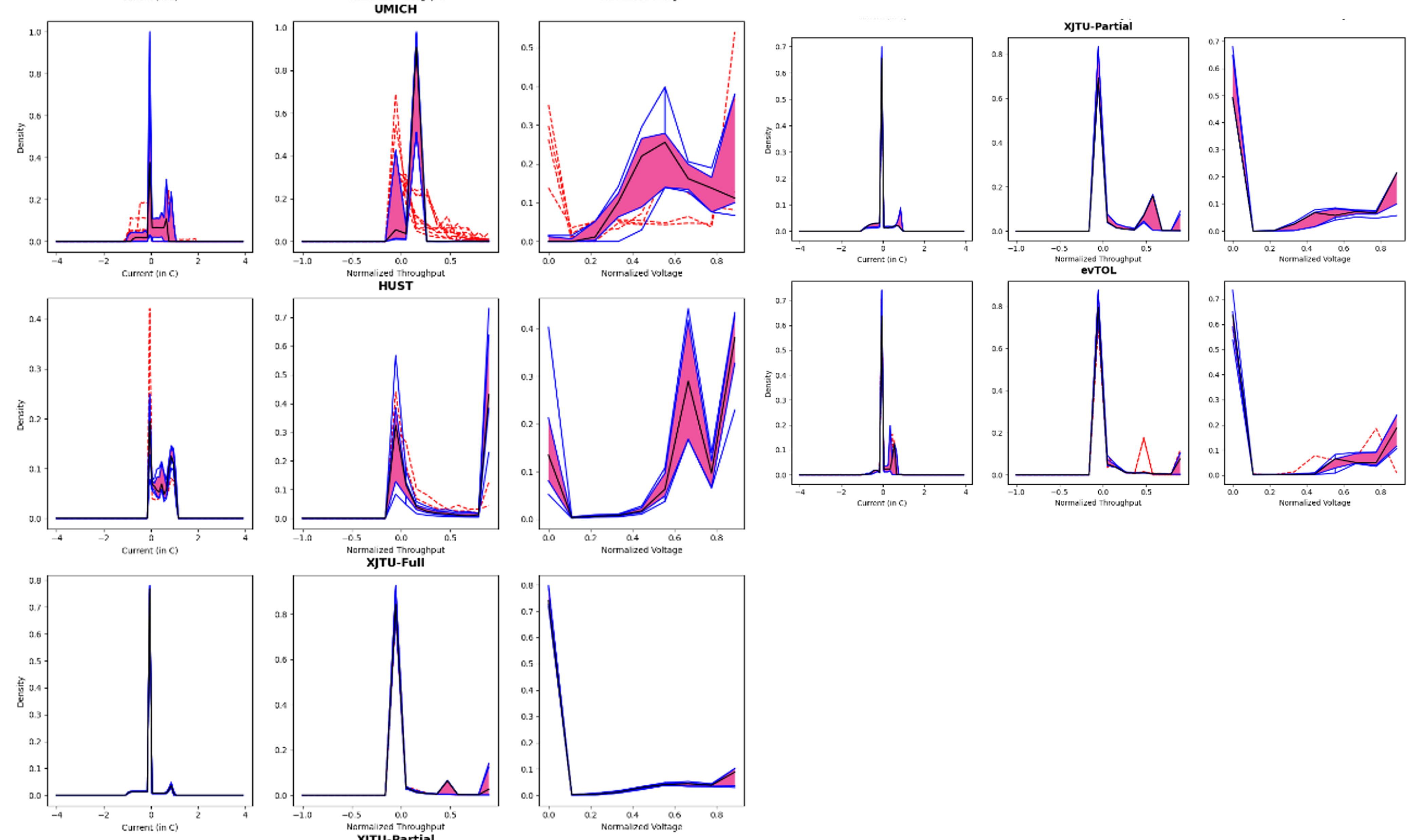}
        \caption{Functional box plots demonstrating the differences between the distributions of current, throughput and voltage in each of the battery datasets used in this work.}
        \label{fig:fb_2}
\end{figure*}

\begin{figure}[t]
  \vskip 0.2in
  \centering
  \centerline{\includegraphics[width=\textwidth]{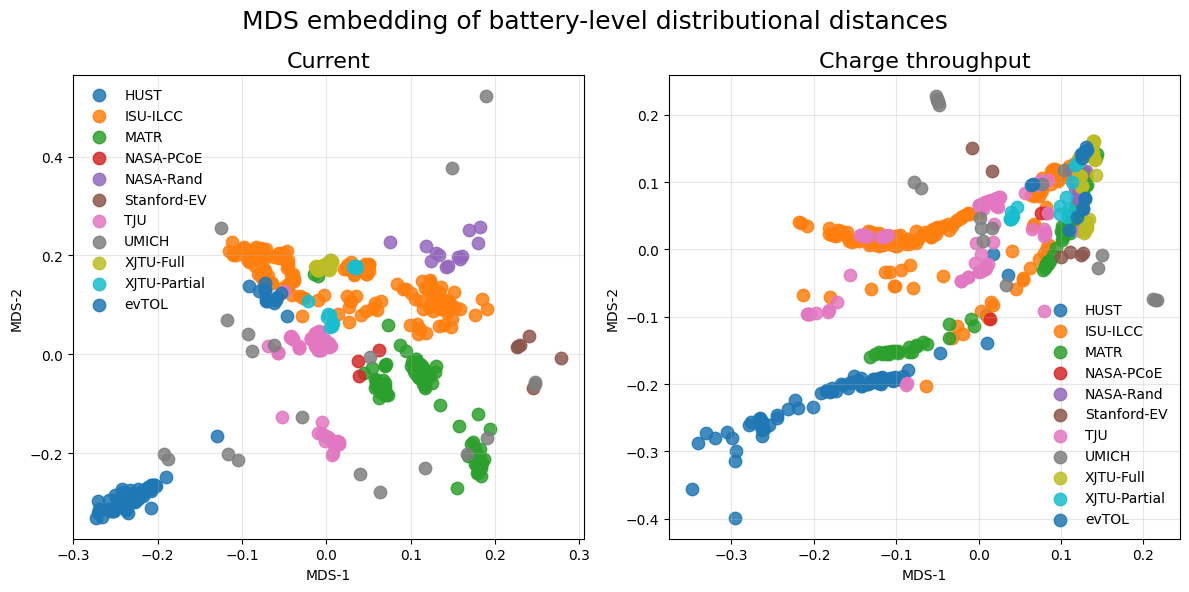}}
    \caption{
      MDS embedding of battery-level distributional distances
    }
    \label{fig:mds}
  \centering
\end{figure}

Figures \ref{fig:fb_plots} and \ref{fig:fb_2} present functional box plot representations of the current, throughput and voltage distributions for each of these datasets to highlight the differences between the operational profiles found in them.
Together, these datasets encompass a broad range of chemistries (NMC, LFP, hybrid), operational depths (partial and full cycling), and profile complexities, making them ideal for assessing the generalizability of the proposed framework across diverse real-world battery operating conditions. Figure \ref{fig:mds}, depicts a multi-dimensional scaling plot visualizing the spread in distributions of current and charge throughput in samples from these datasets.

\section{Baselines}\label{app:baselines}
To comprehensively evaluate the performance of \texttt{NOCE-Net}, we benchmarked it against several recent state-of-the-art sequence-to-sequence modeling approaches that have demonstrated strong performance in time-series forecasting tasks. These baselines cover a diverse spectrum of architectural paradigms—from transformer-based models to state-space models and lightweight MLP-based temporal mixers.

\textbf{TimerXL} \cite{liu2025timerxl} represents a foundational transformer-based framework for universal long-context time series forecasting. It applies a standard transformer architecture to flattened time series representations. However, despite its generality, TimerXL lacks explicit mechanisms to handle variable-length sequences and incurs substantial computational overhead during training and inference, particularly on long lifecycle datasets such as those encountered in battery aging studies.

\textbf{MAMBA} \cite{gu2024mamba} is a selective state-space model developed at Stanford University, which integrates the recurrence efficiency of linear dynamical systems with input-dependent parameterization. Its selective scanning mechanism allows it to effectively handle variable-length input sequences while maintaining linear computational complexity with respect to sequence length. This makes it a suitable candidate for modeling long and irregular battery degradation trajectories.

\textbf{AutoTimes} \cite{NEURIPS2024_dcf88cbc} leverages a pretrained GPT-2 backbone to enable autoregressive time series forecasting via large language model (LLM) adaptation. It fine-tunes language model representations for temporal pattern learning but remains computationally expensive and susceptible to overfitting on small datasets without strong inductive biases.

\textbf{TinyTimeMixers} (TTM) \cite{ekambaram2024tiny} adopts a radically different approach—replacing heavy attention blocks with lightweight MLP-based temporal mixers. These mixers combine multiple small MLPs that operate over short temporal windows, achieving an excellent trade-off between computational efficiency and modeling expressivity.

\begin{table*}[ht]
\centering
\caption{Parameter count comparison between baselines and \texttt{NOCE-Net}.
M1~=~Method~1, M2~=~Method~2, Indep.~=~Independent.}
\label{tab:param_comp}
\small
\setlength{\tabcolsep}{5pt}
\renewcommand{\arraystretch}{1.15}
\resizebox{\textwidth}{!}{%
\begin{tabular}{ccc ccc cc cc c}
\toprule
\multicolumn{3}{c}{\textbf{TimerXL}}
  & \multicolumn{3}{c}{\textbf{AutoTimes}}
  & \multicolumn{2}{c}{\textbf{TTM}}
  & \multicolumn{2}{c}{\textbf{MAMBA}}
  & \textbf{\texttt{NOCE-Net}} \\
\cmidrule(lr){1-3}\cmidrule(lr){4-6}\cmidrule(lr){7-8}
\cmidrule(lr){9-10}\cmidrule(lr){11-11}
\textbf{M1} & \textbf{M2} & \textbf{Indep.}
  & \textbf{M1} & \textbf{M2} & \textbf{Indep.}
  & \textbf{M1} & \textbf{Indep.}
  & \textbf{M2} & \textbf{Indep.}
  & \\
\midrule
269M & 40M & 12M
  & 128M & 124M & 12M
  & 194M & 11M
  & 12M & 7M
  & \textbf{2.5M} \\
\bottomrule
\end{tabular}}
\end{table*}

Each baseline was trained and evaluated under three distinct experimental modes to assess their robustness and adaptability to varying temporal structures:

\begin{table}[t]
\centering
\caption{Structural comparison of the baseline input-representation modes against NOCE-Net.}
\label{tab:mode_comparison}
\small
\resizebox{\textwidth}{!}{
\begin{tabular}{lcccl}
\toprule
\textbf{Mode} & \textbf{Nested structure} & \textbf{Intra-cycle embedding} & \textbf{Variable length} & \textbf{Inter-cycle memory} \\
\midrule
M1 (F, F) & \cmark & MLP                & Pooled to fixed length  & \cmark (baseline's own) \\
M1 (F, T) & \cmark & MLP                & Preserved (padded)      & \cmark \\
M1 (T, F) & \cmark & Multi-patch conv.  & Pooled to fixed length  & \cmark \\
M1 (T, T) & \cmark & Multi-patch conv.  & Preserved (padded)      & \cmark \\
M2        & \xmark & Flat single-level  & N/A                     & Implicit only \\
Indep.    & \xmark & Flat single-level  & N/A                     & None \\
\midrule
\textbf{NOCE-Net} & \textbf{\cmark} & \textbf{Multi-patch conv.} & \textbf{Preserved natively} & \textbf{\cmark~GRU + task/FEC conditioning} \\
\bottomrule
\end{tabular}}
\end{table}

\begin{enumerate}
    \item \textbf{M1: Nested Operations}: It retains the nested operational context structure along with two binary controls—MPATCH and VARSEQ. Intra-cycle input is embedded using a multi-patch embedding layer when MPATCH = True, and an MLP when MPATCH = False. The ORL is used for output reconstruction when MPATCH = True. When VARSEQ = False, the variable intra-cycle length is converted to a fixed length using attention pooling, otherwise a max length padding is applied to the data to fix the sequence length. During voltage prediction, an attention unpooling mechanism is applied to convert to variable length voltage responses when VARSEQ is False. This configuration yields four sub-experiments per baseline.
    \item \textbf{M2: Flattened Operations}: Models are applied to the flattened nested operational context data.
    \item \textbf{Indep. : Independent Mode} : Each intra-cycle input is independently mapped to the corresponding output.
\end{enumerate}

\section{Experimental Details}\label{app:expdetails}

\subsection{Loss Function}\label{app:lossfn}
For a particular FEC, $c$, the training loss is defined as
\begin {align}
    & \mathcal{L} = \mathcal{L}_{MSE} + \alpha * \mathcal{L}_{Diff} + \beta * \mathcal{L}_{Q} \\
    & \mathcal{L}_{MSE}(\mathbf{V}_c, \hat{\mathbf{V}}_c) = ||\mathbf{V}_c - \hat{\mathbf{V}}||_2^2\\
    & \mathcal{L}_{Diff} = \frac{1}{L_c}\sum_{i=2}^{L_c} |\frac{\Delta \mathbf{V}[i]}{\Delta \mathbf{t}[i]} - \frac{\Delta \hat{\mathbf{V}}[i]}{\Delta \mathbf{t}[i]}|\\
    & \mathcal{L}_{Q} = \frac{\sum_i |\mathbf{V}[i]-\hat{\mathbf{V}}[i]| \odot \mathbb{I}[|\mathbf{V}[i]-\hat{\mathbf{V}}[i]|> q_{75}]}{\sum \mathbb{I}[|\mathbf{V}[i]-\hat{\mathbf{V}}[i]|> q_{75}]}
\end{align}
where $q_{75}$ is the $75\%$-ile absolute error value and $\alpha$ and $\beta$ are fixed constants.

\subsection{Architectural Configurations}\label{app:architecture}

\begin{table}[t]
\caption{NOCE-Net model and training hyperparameters. Final values
after all configuration overrides.}
\label{tab:hyperparameters}
\vspace{3pt}
\centering
\small
\setlength{\tabcolsep}{6pt}
\renewcommand{\arraystretch}{1.08}
\begin{tabular}{l l r}
\toprule
\textbf{Group} & \textbf{Hyperparameter} & \textbf{Value} \\
\midrule
\multirow{7}{*}{\textit{Patch Encoder}}
  & Input dimension $d_u$           & 6 \\
  & Model dimension $d_\text{model}$ & 64 \\
  & Patch sizes $\{s_k\}$           & \{4, 16, 32\} \\
  & Number of patches $K$           & 3 \\
  & Encoder Transformer layers      & 6 \\
  & Attention heads                 & 8 \\
  & FFN width multiplier            & 2 \\
\midrule
\multirow{3}{*}{\textit{GRU Backbone}}
  & GRU layers                      & 2 \\
  & Task embedding dim $d_\eta$     & 6 \\
  & FEC embedding dim $d_\gamma$    & 2 \\
\midrule
\multirow{2}{*}{\textit{Patch Decoder}}
  & Decoder Transformer layers      & 3 \\
  & Output dimension $d_v$          & 1 \\
\midrule
\multirow{6}{*}{\textit{Training}}
  & Attention dropout               & 0.1 \\
  & FFN dropout                     & 0.1 \\
  & Cycle batch size                & 2 \\
  & Loss weight $\alpha$ ($\mathcal{L}_\text{Diff}$)  & 0.05 \\
  & Loss weight $\beta$ ($\mathcal{L}_Q$)             & 0.2 \\
  & Max FEC index                   & 5000 \\
\bottomrule
\end{tabular}
\end{table}
The complete list of architectural and training configuration parameters are given in Table \ref{tab:hyperparameters}.
All the transformer blocks in NOCE-Net are implemented with query-key normalization \cite{henry-etal-2020-query}, pre-normalization with SwiGLU activation\cite{xiong2020layernormalizationtransformerarchitecture}, xPoS embeddings\cite{sun2023retentivenetworksuccessortransformer}, and multi-query attention\cite{shazeer2019fasttransformerdecodingwritehead} following current best practices.
The \texttt{Task-Emb} models consist of a two layer MLP with the first layer consisting of a mix of categorical embeddings for categorical features like anode, cathode, form factor, etc., and a linear projection layer for continuous variables like voltage limits and nominal capacity. Each individual feature is embedded to the $d_{model}$ dimension and then concatenated. The second layer is a linear projection layer, $\texttt{Dense}(out =d_{model}, in =d_\eta *d_{model})$. The \texttt{FEC-Emb} model is a single linear projection.
The model was trained end to end using the Adam optimizer and a step scheduler (step size = $5$, gamma = $0.33$) with an initial learning rate of $1e-2$. 

\subsection{Training Configurations}\label{app:trainconfigs}

Each sample in a minibatch used to train  
\texttt{NOCE-Net} consists of battery operational inputs till end of life in a nested formulation, which leads to a very high dimensional data sample. To fit the process in compute memory, the total life of the battery is split into consecutive non-overlapping groups of 100 cycles. In each training step, corresponding to a data sample, the model is trained sequentially on these groups of 100 cycles. The recurrent states generated after each collection of 100 cycles is detached from the computational graph and then passed as the initial state for the next 100 cycles. The remaining computation happens parallely. 
For stateless baselines using Method 1 and 2, like TimerXL, AutoTimes and TTM, each sample is broken down into non-overlapping consecutive sequences of 100 cycles and each such sequence is treated as an independent sample. Being a recurrent architecture, MAMBA is trained similar to \texttt{NOCE-Net} while allowing the flow of recurrent states with detachment after every 100 cycles.

\section{Detailed Results for Voltage Forecasting}\label{app:forecastres}

 \subsection{In-sample Prediction}\label{app:insampleres}
 Table \ref{tab:in-sample zero-shot} shows the results of in-sample prediction on the test split of the training dataset using the trained models.

\begin{table*}[ht]
\centering
\caption{In-sample prediction results across datasets and aggregate.
Values are the average of 5 independent runs.
\textbf{Bold}~=~best, \underline{underline}~=~second best.}
\label{tab:in-sample zero-shot}
\scriptsize
\setlength{\tabcolsep}{3pt}
\renewcommand{\arraystretch}{1.15}
\resizebox{\textwidth}{!}{%
\begin{tabular}{llllcccccccccccccc}
\toprule
 &  &  &  &
 \multicolumn{2}{c}{\textbf{ISU-ILCC}} &
 \multicolumn{2}{c}{\textbf{NASA-Rand}} &
 \multicolumn{2}{c}{\textbf{TJU}} &
 \multicolumn{2}{c}{\textbf{MATR}} &
 \multicolumn{2}{c}{\textbf{UMICH-Pouch}} &
 \multicolumn{2}{c}{\textbf{XJTU-Partial}} &
 \multicolumn{2}{c}{\textbf{Total}} \\
\cmidrule(lr){5-6}\cmidrule(lr){7-8}\cmidrule(lr){9-10}
\cmidrule(lr){11-12}\cmidrule(lr){13-14}\cmidrule(lr){15-16}\cmidrule(lr){17-18}
\textbf{Method} & \textbf{VS} & \textbf{MP} & \textbf{Model}
 & MAE & MAPE & MAE & MAPE & MAE & MAPE
 & MAE & MAPE & MAE & MAPE & MAE & MAPE
 & MAE & MAPE \\
\midrule

\multirow{12}{*}{\textbf{M1}}
 & F & F & \multirow{4}{*}{\textbf{AutoTimes}}
 & 0.1185 & 16.310 & 0.1899 & 37.194 & 0.3184 & 32.571
 & 0.0452 & 10.641 & 0.1516 & 23.809 & 0.1120 & 18.398
 & 0.1598 & 20.709 \\
 & F & T &
 & 0.1003 & 13.684 & 0.2257 & 42.557 & 0.3038 & 28.880
 & 0.0492 & 13.351 & 0.1753 & 27.385 & 0.0709 & 11.261
 & 0.1470 & 18.634 \\
 & T & F &
 & 0.1141 & 15.166 & 0.1932 & 38.845 & 0.3750 & 45.488
 & 0.1532 & 60.372 & 0.1638 & 26.662 & 0.1096 & 18.438
 & 0.1802 & 27.317 \\
 & T & T &
 & 0.1443 & 18.874 & 0.1852 & 38.068 & 0.3787 & 46.768
 & 0.2285 & 90.582 & 0.1742 & 28.419 & 0.1437 & 23.061
 & 0.2066 & 32.524 \\
\cmidrule(lr){2-18}
 & F & F & \multirow{4}{*}{\textbf{TTM}}
 & 0.1148 & 15.561 & 0.2060 & 38.617 & 0.3156 & 32.450
 & 0.0435 & 10.895 & 0.1487 & 23.418 & 0.1169 & 18.983
 & 0.1576 & 20.355 \\
 & F & T &
 & \underline{0.0627} & \underline{8.615} & 0.2225 & 39.672 & 0.3089 & \underline{27.685}
 & \underline{0.0376} & \underline{9.883} & \textbf{0.1127} & \textbf{17.348} & 0.0489 & 8.217
 & 0.1229 & \underline{14.633} \\
 & T & F &
 & 0.1362 & 17.846 & 0.1919 & 37.790 & 0.3807 & 45.801
 & 0.1708 & 74.782 & 0.1652 & 26.659 & 0.1277 & 21.093
 & 0.1964 & 30.228 \\
 & T & T &
 & 0.1496 & 19.517 & 0.1930 & 38.766 & 0.3759 & 47.397
 & 0.2222 & 93.385 & 0.1778 & 29.387 & 0.1453 & 24.306
 & 0.2089 & 33.384 \\
\cmidrule(lr){2-18}
 & F & F & \multirow{4}{*}{\textbf{TimerXL}}
 & 0.2249 & 30.723 & 0.3817 & 73.128 & 0.6114 & 62.699
 & 0.0856 & 20.772 & 0.2895 & 45.535 & 0.2207 & 36.057
 & 0.3060 & 39.593 \\
 & F & T &
 & 0.1659 & 22.725 & 0.4687 & 85.907 & 0.6416 & 59.200
 & 0.0896 & 23.920 & 0.2941 & 45.605 & 0.1226 & 20.035
 & 0.2799 & 34.378 \\
 & T & F &
 & 0.1660 & 21.907 & 0.2575 & 51.234 & 0.5045 & 60.949
 & 0.2158 & 89.510 & 0.2197 & 35.609 & 0.1575 & 26.263
 & 0.2506 & 38.279 \\
 & T & T &
 & 0.1606 & 20.984 & 0.2067 & 41.996 & 0.4125 & 51.467
 & 0.2464 & 100.55 & 0.1924 & 31.596 & 0.1580 & 25.891
 & 0.2271 & 36.023 \\
\midrule

\multirow{3}{*}{\textbf{M2}}
 & \multicolumn{3}{l}{\textbf{AutoTimes}}
 & 0.0923 & 12.579 & \underline{0.1621} & \underline{35.856} & 0.3062 & 36.977
 & 0.1556 & 59.858 & 0.1880 & 31.797 & 0.1009 & 16.880
 & 0.1519 & 23.847 \\
 & \multicolumn{3}{l}{\textbf{MAMBA}}
 & 0.0705 & 9.874 & 0.1969 & 41.130 & \textbf{0.2940} & 28.917
 & 0.1371 & 56.072 & 0.1299 & 21.274 & 0.0504 & 8.647
 & 0.1331 & 19.790 \\
 & \multicolumn{3}{l}{\textbf{TimerXL}}
 & 0.1174 & 16.217 & 0.2606 & 55.994 & 0.4389 & 47.590
 & 0.2119 & 84.005 & 0.2297 & 38.220 & 0.1080 & 18.299
 & 0.2066 & 31.550 \\
\midrule

\multirow{3}{*}{\textbf{Indep.}}
 & \multicolumn{3}{l}{\textbf{AutoTimes}}
 & 0.1011 & 13.715 & 0.1775 & 39.266 & 0.3354 & 66.729
 & 0.1705 & 65.654 & 0.2059 & 35.155 & 0.1150 & 18.507
 & 0.1664 & 27.076 \\
 & \multicolumn{3}{l}{\textbf{TTM}}
 & 0.1590 & 20.630 & 0.2828 & 49.383 & 0.4795 & 68.473
 & 0.1630 & 55.049 & 0.2085 & 36.323 & 0.1499 & 26.329
 & 0.2368 & 30.002 \\
 & \multicolumn{3}{l}{\textbf{MAMBA}}
 & 0.0635 & 8.960 & 0.1774 & 37.186 & 0.2980 & 29.241
 & 0.1235 & 50.369 & \underline{0.1170} & \underline{19.152} & \underline{0.0454} & \underline{7.689}
 & \underline{0.1199} & 16.821 \\
\midrule

\multicolumn{4}{l}{\textbf{\texttt{NOCE-Net} (Proposed)}}
 & \textbf{0.0260} & \textbf{4.038}
 & \textbf{0.0624} & \textbf{13.888}
 & \underline{0.2950} & \textbf{25.584}
 & \textbf{0.0250} & \textbf{8.927}
 & 0.1456 & 23.866
 & \textbf{0.0378} & \textbf{6.227}
 & \textbf{0.0914} & \textbf{10.383} \\
\bottomrule
\end{tabular}}
\vspace{0.4em}
\raggedright\scriptsize
\textbf{VS}~=~VARSEQ, \textbf{MP}~=~MPATCH, \textbf{F}~=~False,
\textbf{T}~=~True, \textbf{M1}~=~Method~1, \textbf{M2}~=~Method~2,
\textbf{Indep.}~=~Independent
\end{table*}

\subsection{Few Shot Forecasting}\label{app:fewshotres}
To assess generalization in limited-data regimes, we conducted few-shot forecasting experiments, where the final projection heads of all models were fine-tuned using the first 50 FECs of each test cell, while freezing all other parameters. Baselines underwent 50 steps of gradient updates each and \texttt{NOCE-Net} underwent 10. Additionally, \texttt{NOCE-Net} was also finetuned with 100 initial FEC data. As shown in Table \ref{tab:few-shot}, finetuning significantly reduces the errors across all the baselines. Specifically, MAMBA + Method 2 performs best in three datasets outperforming \texttt{NOCE-Net}. However, \texttt{NOCE-Net} outperforms the baselines in the aggregate setting, outperforming MAMBA + Method 2 by $42.5\%$ and TTM + Method-1(False, True) by $30\%$ in terms of MAE. Interestingly, when finetuned with 100 cycles, \texttt{NOCE-Net} performs worse compared to that of 50 cycles. This possibly stems from overfitting of the limited projection parameters with an increased data size.

\begin{table*}[t]
\caption{Few-shot learning results on the test dataset. Values are the
average of 5 independent runs (std.\ $3$--$7\%$; higher when
\textsc{VarSeq}=True). \textbf{Bold}~=~best,
\underline{underline}~=~second best.}
\label{tab:few-shot}
\centering
\scriptsize
\setlength{\tabcolsep}{3pt}
\renewcommand{\arraystretch}{1.15}
\resizebox{\linewidth}{!}{%
\begin{tabular}{llllcccccccccccc}
\toprule
 & & & &
 \multicolumn{2}{c}{\textbf{NASA-PCoE}} &
 \multicolumn{2}{c}{\textbf{Stanford-EV}} &
 \multicolumn{2}{c}{\textbf{HUST}} &
 \multicolumn{2}{c}{\textbf{XJTU-Full}} &
 \multicolumn{2}{c}{\textbf{evTOL}} &
 \multicolumn{2}{c}{\textbf{Total}} \\
\cmidrule(lr){5-6}\cmidrule(lr){7-8}\cmidrule(lr){9-10}
\cmidrule(lr){11-12}\cmidrule(lr){13-14}\cmidrule(lr){15-16}
\textbf{Method} & \textbf{VS} & \textbf{MP} & \textbf{Model}
 & MAE & MAPE & MAE & MAPE & MAE & MAPE
 & MAE & MAPE & MAE & MAPE & MAE & MAPE \\
\midrule

\multirow{8}{*}{\textbf{M1}}
 & F & F & \multirow{4}{*}{\textbf{AutoTimes}}
 & 0.1719 & 29.220 & 0.1226 & 15.782 & 0.1335 & 39.631
 & 0.1035 & 18.325 & 0.1671 & 26.711 & 0.1328 & 33.241 \\
 & F & T &
 & 0.1458 & 23.721 & \underline{0.0781} & \underline{10.113} & 0.0669 & 45.122
 & 0.0472 & 8.518 & 0.1057 & 17.308 & 0.0702 & 33.909 \\
 & T & F &
 & 0.1737 & 27.785 & 0.1112 & 15.666 & 0.1317 & 81.319
 & 0.1472 & 28.255 & 0.1554 & 23.009 & 0.1355 & 61.624 \\
 & T & T &
 & 0.1792 & 30.729 & 0.1208 & 15.843 & 0.1118 & 74.682
 & 0.1736 & 29.739 & 0.1343 & 23.352 & 0.1243 & 57.538 \\
\cmidrule(lr){2-16}
 & F & F & \multirow{4}{*}{\textbf{TTM}}
 & 0.1705 & 29.081 & 0.1100 & 14.494 & 0.1225 & 38.319
 & 0.1085 & 17.528 & 0.1690 & 27.264 & 0.1255 & 32.231 \\
 & F & T &
 & 0.0880 & 14.555 & 0.0790 & 10.218 & 0.0553 & 43.056
 & \underline{0.0324} & \underline{6.022} & 0.0983 & 15.183 & 0.0593 & 31.880 \\
 & T & F &
 & 0.2015 & 30.802 & 0.1169 & 15.426 & 0.1540 & 71.465
 & 0.1876 & 35.812 & 0.1409 & 22.904 & 0.1547 & 56.181 \\
 & T & T &
 & 0.1956 & 32.496 & 0.1208 & 15.860 & 0.0885 & 61.986
 & 0.1737 & 29.780 & 0.1367 & 23.289 & 0.1094 & 49.192 \\
\midrule

\multirow{2}{*}{\textbf{M2}}
 & \multicolumn{3}{l}{\textbf{AutoTimes}}
 & 0.1117 & 19.800 & 0.1401 & 18.659 & 0.1205 & 62.884
 & 0.0975 & 15.910 & 0.1047 & 19.239 & 0.1168 & 46.299 \\
 & \multicolumn{3}{l}{\textbf{MAMBA}}
 & \textbf{0.0480} & \textbf{9.930} & \textbf{0.0494} & \textbf{6.453} & 0.0913 & 66.662
 & \textbf{0.0265} & \textbf{5.151} & 0.0464 & 9.815 & 0.0721 & 44.781 \\
\midrule

\multicolumn{4}{l}{\textbf{\texttt{NOCE-Net}: 50-cycle shot}}
 & \underline{0.0669} & \underline{12.089} & 0.1108 & 13.501
 & \textbf{0.0326} & \textbf{14.117}
 & 0.0377 & 7.713 & \textbf{0.0450} & \underline{9.427}
 & \textbf{0.0414} & \textbf{8.233} \\
\multicolumn{4}{l}{\textbf{\texttt{NOCE-Net}: 100-cycle shot}}
 & 0.0716 & 13.310 & 0.0938 & 11.491
 & \underline{0.0343} & \underline{15.260}
 & 0.0399 & 7.739 & \underline{0.0456} & \textbf{9.336}
 & \underline{0.0407} & \underline{8.370} \\
\bottomrule
\end{tabular}}
\vspace{0.4em}
\raggedright\scriptsize
\textbf{VS}~=~VARSEQ, \textbf{MP}~=~MPATCH, \textbf{F}~=~False,
\textbf{T}~=~True, \textbf{M1}~=~Method~1, \textbf{M2}~=~Method~2
\end{table*}

\section{SOH Transferability: Extended Analysis}
\label{app:soh_analysis}

This appendix provides full numerical results, architectural details, and
extended discussion for Section~\ref{sec:res_soh_transferability}.

\subsection{Spearman Rank Correlation Diagnostics}
\label{app:spearman}

\begin{table}[t]
\caption{Spearman rank correlation ($|\rho|$) diagnostics for NOCE-Net GRU hidden
states across all dataset splits. Left block (\textit{Repr.\ vs.\ cycle index}):
monotonicity of four scalar summaries of $h_c$ against FEC index $c$. Right block
(\textit{Capacity vs.\ repr.}): alignment of normalized capacity trajectory $Q(c)$ with each
summary. \textbf{Bold}: best per row within each block. \textit{All} rows report
mean $|\rho|$ over all batteries in the split.}
\label{tab:monotonicity_diagnostics}
\centering
\setlength{\tabcolsep}{4pt}
\renewcommand{\arraystretch}{1.08}
\resizebox{\columnwidth}{!}{%
\small
\begin{tabular}{%
  l
  S[table-format=1.3]
  S[table-format=1.3]
  S[table-format=1.3]
  S[table-format=1.3]
  @{\hspace{6pt}}
  S[table-format=1.3]
  S[table-format=1.3]
  S[table-format=1.3]
  S[table-format=1.3]
}
\toprule
& \multicolumn{4}{c}{\textit{Repr.\ vs.\ cycle index} $|\rho(\cdot,\, c)|$}
& \multicolumn{4}{c}{\textit{Capacity vs.\ repr.} $|\rho(Q,\, \cdot)|$} \\
\cmidrule(lr){2-5}\cmidrule(lr){6-9}
{Dataset}
  & {$\|h_c\|$}
  & {$\mathrm{PC1}(H)$}
  & {$\|h_c{-}h_0\|$}
  & {$\mathrm{PC1}(\Delta H)$}
  & {$\|h_c\|$}
  & {$\mathrm{PC1}(H)$}
  & {$\|h_c{-}h_0\|$}
  & {$\mathrm{PC1}(\Delta H)$} \\
\midrule

\multicolumn{9}{l}{\textsc{Train}} \\[1pt]
ISU\_ILCC    & 0.919 & \textbf{0.999} & 0.998 & \textbf{0.999}
             & 0.919 & \textbf{0.999} & 0.998 & \textbf{0.999} \\
NASA Rand    & \textbf{0.970} & 0.997 & 0.974 & 0.996
             & 0.921 & 0.943 & 0.924 & \textbf{0.945} \\
TJU      & 0.993 & \textbf{1.000} & 0.999 & \textbf{1.000}
             & 0.993 & \textbf{1.000} & 0.999 & \textbf{1.000} \\
MATR     & 0.525 & \textbf{1.000} & \textbf{1.000} & \textbf{1.000}
             & 0.522 & \textbf{0.995} & \textbf{0.995} & \textbf{0.995} \\
XJTU-Partial & 0.996 & \textbf{1.000} & 0.999 & \textbf{1.000}
             & 0.857 & 0.848 & 0.854 & \textbf{0.855} \\
\rowcolor{rowgray}
\textit{All} & 0.825 & \textbf{1.000} & 0.998 & \textbf{1.000}
             & 0.808 & \textbf{0.990} & 0.989 & \textbf{0.991} \\

\midrule
\multicolumn{9}{l}{\textsc{Validation}} \\[1pt]
ISU\_ILCC    & 0.902 & \textbf{1.000} & 0.999 & \textbf{1.000}
             & 0.902 & \textbf{1.000} & 0.999 & \textbf{1.000} \\
NASA Rand    & 0.818 & 0.995 & 0.949 & \textbf{0.996}
             & 0.817 & \textbf{0.989} & 0.941 & \textbf{0.989} \\
TJU      & 0.999 & \textbf{1.000} & \textbf{1.000} & \textbf{1.000}
             & 0.999 & \textbf{1.000} & \textbf{1.000} & \textbf{1.000} \\
MATR     & 0.333 & \textbf{1.000} & \textbf{1.000} & \textbf{1.000}
             & 0.334 & \textbf{0.999} & \textbf{0.999} & \textbf{0.999} \\
XJTU-Partial & 0.999 & \textbf{1.000} & \textbf{1.000} & \textbf{1.000}
             & 0.982 & 0.982 & 0.983 & \textbf{0.983} \\
\rowcolor{rowgray}
\textit{All} & 0.762 & \textbf{1.000} & 0.998 & \textbf{1.000}
             & 0.749 & \textbf{0.998} & 0.996 & \textbf{0.998} \\

\midrule
\multicolumn{9}{l}{\textsc{Test (Zero-Shot)}} \\[1pt]
NASA PCoE    & \textbf{0.989} & \textbf{1.000} & 0.996 & \textbf{1.000}
             & 0.686 & 0.668 & \textbf{0.688} & 0.668 \\
Stanford-EV  & 0.967 & \textbf{0.997} & 0.980 & \textbf{0.997}
             & 0.960 & \textbf{0.986} & 0.967 & \textbf{0.986} \\
HUST         & 0.612 & \textbf{1.000} & \textbf{1.000} & \textbf{1.000}
             & 0.613 & \textbf{1.000} & \textbf{1.000} & \textbf{1.000} \\
XJTU-Full    & 0.996 & \textbf{1.000} & \textbf{1.000} & \textbf{1.000}
             & 0.852 & 0.839 & 0.850 & \textbf{0.850} \\
evTOL$^*$    & 0.673 & \textbf{0.997} & 0.982 & \textbf{0.997}
             & 0.235 & 0.277 & 0.275 & \textbf{0.289} \\
\rowcolor{rowgray}
\textit{All} & 0.750 & \textbf{1.000} & 0.997 & \textbf{1.000}
             & 0.659 & \textbf{0.874} & 0.877 & \textbf{0.879} \\
\bottomrule
\end{tabular}%
}
\vspace{3pt}
\begin{minipage}{\columnwidth}
  \footnotesize
  $^*$evTOL batteries exhibit non-monotone capacity trajectories due to
  aircraft partial-discharge profiles; low $|\rho(Q,\cdot)|$ reflects target
  signal structure rather than representation quality
  ($|\rho(\mathrm{PC1},c)| = 0.997$).
\end{minipage}
\end{table}

\begin{figure}[ht]
  \centering
  \includegraphics[width=\linewidth]{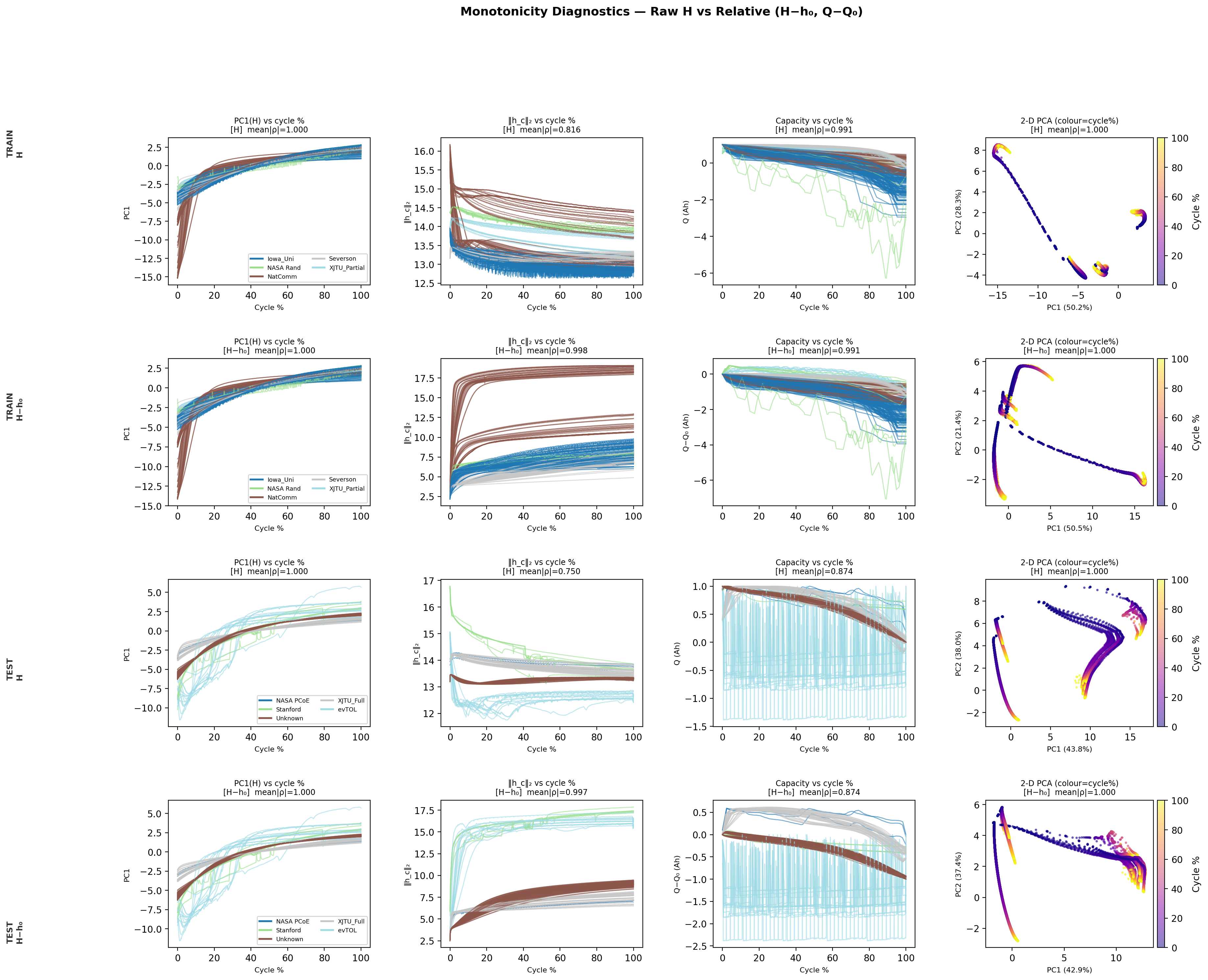}
  \caption{%
    Full monotonicity diagnostic grid.
    Rows: train split with raw $H$ and $H - h_0$ (top two), test split
    with raw $H$ and $H - h_0$ (bottom two).
    Columns (left to right): PC1 score vs.\ cycle\%, L2 norm vs.\ cycle\%,
    capacity vs.\ cycle\%, and 2-D PCA trajectory coloured by cycle\%.
    Mean $|\rho|$ values are annotated in each panel title.
    There are some differently named legend titles which are clarified below:
    Iowa\_Uni = ISU\_ILCC, Unknown = HUST, Severson = MATR, NatComm = TJU}
  \label{fig:full_diagnostics}
\end{figure}

Table~\ref{tab:monotonicity_diagnostics} reports Spearman $|\rho|$ between
the FEC index and four scalar summaries of the GRU hidden states, as well as
the correlation between each summary and the capacity trajectory~$Q(c)$.
Two findings merit emphasis beyond what is reported in the main text.
First, $|\rho|_{Q}$, the spearman correlation of normalized capacity fade (SOH) are close to one, confirming that capacity fade is strictly monotone at the FEC
timescale, for all well-behaved datasets (except NASA and evTOL).
Second, the norm $\|h_c\|_2$ shows a large drop in $|\rho|$ specifically for
MATR on validation ($0.333$) and HUST on test ($0.612$), while
$\mathrm{PC1}(H)$ maintains $|\rho| = 1.000$ for both.
Both datasets contain large batteries-per-dataset counts with diverse
cycling protocols, producing high inter-battery variance in $\|h_0\|_2$
that possibly corrupts the aggregate rank ordering even though within-battery
monotonicity is perfect.
This is likely the mechanism by which the relativisation $h_c - h_0$ recovers the norm $|\rho|$ from $0.612$ to $0.997$ for HUST.

\subsection{PID-R Feature Extraction from GRU Hidden States}
\label{app:pid_features}

Let $\{h_c\}_{c=1}^{N} \in \mathbb{R}^{d_h}$ denote the GRU recurrent
states output by NOCE-Net for a battery with $N$ full equivalent cycles.
To construct a richer set of features for downstream SOH regression, we
extract four complementary temporal summaries of the hidden state
sequence, motivated by the proportional--integral--derivative~(PID)
decomposition commonly used in control theory.
We refer to these collectively as \textbf{PID-R features}.

\paragraph{Proportional (P).}
The raw hidden state at each cycle,
\begin{equation}
  P_c = h_c \in \mathbb{R}^{d_h},
\end{equation}
encodes the model's current assessment of the battery's operational
context at FEC~$c$.

\paragraph{Derivative (D).}
The cycle-to-cycle change in hidden state,
\begin{equation}
  D_c = h_c - h_{c-1} \in \mathbb{R}^{d_h}, \quad D_1 = \mathbf{0},
\end{equation}
captures the instantaneous rate of change of battery dynamics.
Large $\|D_c\|$ indicates a rapid shift in operational context or
degradation state, providing a local degradation-rate signal that is
complementary to the cumulative information in $P_c$.

\paragraph{Integral (I).}
A learnable exponential moving average~(EMA) of the hidden states,
\begin{equation}
  I_c = \lambda \odot I_{c-1} + (\mathbf{1} - \lambda) \odot h_c,
  \quad I_1 = h_1,
\end{equation}
where $\lambda \in \mathbb{R}^{d_h}$ is a learned per-dimension decay
vector constrained to $(0.5,\, 0.999)$ via a sigmoid
reparameterisation.
Unlike a direct cumulative sum, the EMA is bounded and stationary,
with each dimension's effective memory horizon determined by
$(1 - \lambda_i)^{-1}$ cycles.
The integral term accumulates a smoothed history of past operational
context, providing a signal that is robust to individual cycle noise
and sensitive to long-term drift in battery dynamics.

\paragraph{Relative (R).}
The hidden state expressed relative to its battery-specific
initialisation,
\begin{equation}
  R_c = h_c - h_0 \in \mathbb{R}^{d_h},
\end{equation}
where $h_0 = \phi_\text{init}(\eta)$ is the task-conditioned initial
hidden state set by the battery metadata embedding.
As demonstrated in Section~\ref{sec:res_soh_transferability} and
Table~\ref{tab:monotonicity_diagnostics}, the raw norm $\|h_c\|$ is
dominated by the battery-specific magnitude of $h_0$, which is
orthogonal to the degradation direction.
The relative feature removes this initialisation offset, recovering
near-perfect Spearman correlation with the SOH trajectory
($|\rho| = 0.997$ on test, compared to $0.750$ for the raw norm).

\paragraph{Combined feature vector.}
The four components are concatenated along the feature dimension to
form the PID-R feature vector,
\begin{equation}
  \phi_c = \bigl[P_c \;\|\; D_c \;\|\; I_c \;\|\; R_c\bigr]
  \in \mathbb{R}^{4d_h},
\end{equation}
which is then passed to the downstream SOH prediction head.
The concatenation preserves all four temporal perspectives simultaneously:
$P_c$ provides the current state, $D_c$ the local rate of change,
$I_c$ the accumulated history, and $R_c$ the net displacement from
the battery's initialisation point.
Together they provide a compact, fixed-dimensional summary of the
battery's degradation trajectory that is computed entirely from the
frozen NOCE-Net hidden states without any additional model training.

For batched inference over variable-length sequences, the EMA
recurrence is computed with a mask that freezes the integral state at
padding positions, ensuring that batteries with different cycle counts
in the same batch do not corrupt each other's integral estimates.
The learnable decay $\lambda$ is initialised to $0.95$ and optimised
jointly with the downstream SOH head.

\subsection{Direct SOH Prediction Head}
\label{app:soh_head}

\begin{table}[t]
\caption{%
  Direct SOH prediction results from the frozen PID-R feature head
  trained on voltage-forecasting representations.
  \textit{In-Sample}: batteries from training datasets held out at
  test time. \textit{Zero-Shot}: unseen datasets and chemistries.
  evTOL excluded from aggregate due to non-monotone capacity profiles
  ($|\rho|_Q = 0.278$); reported separately for completeness.
}
\label{tab:direct_soh}
\vspace{3pt}
\centering
\small
\setlength{\tabcolsep}{6pt}
\renewcommand{\arraystretch}{1.08}
\begin{tabular}{l r S[table-format=3.2] S[table-format=1.4]}
\toprule
\textbf{Dataset} & \textbf{MAPE (\%)} & \textbf{MAE} \\
\midrule
\multicolumn{3}{l}{\textsc{In-Sample Test}} \\[2pt]
MATR              &  39.06 & 0.1284 \\
XJTU-Partial      &  28.59 & 0.0713 \\
TJU               & 102.07 & 0.1941 \\
ISU\_ILCC         & 143.73 & 0.4510 \\
NASA Rand         & 208.97 & 0.6619 \\
\rowcolor{rowgray}
\textit{Total}    &  98.90 & 0.2834 \\
\midrule
\multicolumn{3}{l}{\textsc{Zero-Shot Test}} \\[2pt]
NASA PCoE         &  16.59 & 0.0981 \\
HUST              &  72.75 & 0.1371 \\
XJTU-Full         &  75.78 & 0.1595 \\
Stanford-EV       & 106.11 & 0.7852 \\
evTOL$^*$         &  97.30 & 0.5664 \\
\rowcolor{rowgray}
\textit{Total (excl.\ evTOL)}  & 76.30 & 0.2137 \\
\bottomrule
\end{tabular}
\vspace{3pt}
\begin{minipage}{\columnwidth}
  \footnotesize
  $^*$evTOL exhibits non-monotone capacity.
\end{minipage}
\end{table}

\begin{table}[t]
\caption{%
  FEC-only baseline comparison for SOH prediction, addressing the
  concern that hidden-state/capacity correlation may merely reflect
  cycle progression.
  \textit{PID MAPE}: the PID-R feature + frozen-hidden-state SOH head
  pipeline (full diagnostic in Table~\ref{tab:shape_diagnostic}).
  \textit{Lin/MLP MAPE}: a Linear / 3-layer MLP model mapping only the
  normalised FEC index (no hidden state) to SOH.
  $\Delta = (\text{FEC-only MAPE} - \text{PID MAPE})/\text{PID MAPE}$;
  positive $\Delta$ indicates the FEC-only baseline is worse than the
  PID-R representation.
  \textbf{Bold}: $\Delta < 0$ (FEC-only baseline outperforms PID-R;
  reflects known affine miscalibration of the SOH head on these
  batteries, Table~\ref{tab:shape_diagnostic}).
  evTOL excluded (non-monotone capacity).
}
\label{tab:fec_baseline}
\vspace{3pt}
\centering
\setlength{\tabcolsep}{4pt}
\renewcommand{\arraystretch}{1.08}
\resizebox{\columnwidth}{!}{%
\small
\begin{tabular}{l r r r r r}
\toprule
\textbf{Dataset}
  & \textbf{PID MAPE $\downarrow$}
  & \textbf{Lin MAPE $\downarrow$}
  & \textbf{$\Delta$ Lin}
  & \textbf{MLP MAPE $\downarrow$}
  & \textbf{$\Delta$ MLP} \\
\midrule
\multicolumn{6}{l}{\textsc{In-Sample Test}} \\[2pt]
ISU\_ILCC    & 143.73\% & \textbf{73.69\%}  & \textbf{$-48.7\%$}  & \textbf{88.43\%}  & \textbf{$-38.5\%$} \\
NASA Rand    & 208.97\% & \textbf{106.79\%} & \textbf{$-48.9\%$}  & \textbf{119.00\%} & \textbf{$-43.1\%$} \\
TJU          & 102.07\% & 124.89\%          & $+22.4\%$           & 158.87\%          & $+55.6\%$ \\
MATR         &  39.06\% &  93.18\%          & $+138.6\%$          & 108.69\%          & $+178.3\%$ \\
XJTU-Partial &  28.59\% &  98.90\%          & $+245.9\%$          & 117.11\%          & $+309.6\%$ \\
\rowcolor{rowgray}
\textit{Total} & 98.90\% & 86.01\% & $-13.0\%$ & 103.16\% & $+4.3\%$ \\
\midrule
\multicolumn{6}{l}{\textsc{Zero-Shot Test}} \\[2pt]
NASA PCoE    &  16.59\% &  73.45\%          & $+342.7\%$          &  74.31\%          & $+347.9\%$ \\
Stanford-EV  & 106.11\% & \textbf{68.90\%}  & \textbf{$-35.1\%$}  & \textbf{67.52\%}  & \textbf{$-36.4\%$} \\
HUST         &  72.75\% & 118.04\%          & $+62.3\%$           & 147.92\%          & $+103.3\%$ \\
XJTU-Full    &  75.78\% & 110.32\%          & $+45.6\%$           & 135.01\%          & $+78.2\%$ \\
\rowcolor{rowgray}
\textit{Total} & 76.30\% & 116.51\% & $+52.7\%$ & 145.38\% & $+90.5\%$ \\
\bottomrule
\end{tabular}}
\end{table}

The SOH head takes the PID-R features (see \ref{app:pid_features})
extracted from the frozen GRU hidden states as input and predicts the
normalised SOH trajectory directly via a three-layer MLP with GELU
activations and LayerNorm, followed by a linear output.
The target is the normalised capacity
$\mathrm{SOH}(c) = (Q(c) - 0.8 Q_\text{nom}) / (Q_\text{nom} - 0.8 Q_\text{nom})$
which is $1.0$ at BOL and $0.0$ at the conventional 80\% capacity
threshold, and can become negative for batteries cycled beyond EOL.
The head is trained with a composite loss comprising MSE, soft monotone
penalty, and a BOL anchor term that penalises deviation of $\hat{SOH}(0)$ from
the known initial value $\mathrm{SOH}(0) \approx 1.0$.
No sigmoid activation is applied: the linear output allows unrestricted
prediction of negative SOH for deeply degraded batteries.
UMICH batteries are excluded as their public release does not contain
capacity measurements. evTOL batteries are excluded from capacity metrics
as their aircraft discharge profiles produce non-monotone capacity
trajectories ($|\rho|_Q = 0.278$) that are structurally incompatible with
the monotone SOH formulation; their representation quality
($|\rho|_{\mathrm{PC1}} = 0.997$) remains strong. The SOH head trained jointly on normalized capacity labels from the training dataset provides the base model for the analysis in the later sections.

\subsection{Shape Diagnostic}
\label{app:shape_diagnostic}


\begin{table}[t]
\caption{%
  Shape diagnostic for the direct SOH prediction head.
  \textit{Shape}~$r$: Pearson correlation between centred predicted
  and true SOH trajectories — measures whether the model captures
  the \emph{shape} of capacity fade independently of scale and offset.
  \textit{Offset}~$\bar{\delta} = \overline{\hat{y}} - \bar{y}$:
  signed mean prediction bias.
  \textit{Scale needed}~$\sigma_y/\sigma_{\hat{y}}$: ratio of true
  to predicted trajectory standard deviation; values $>1$ indicate
  a compressed predicted fade amplitude.
  \textbf{Bold}: shape $r < 0.80$.
  evTOL excluded from aggregate.
}
\label{tab:shape_diagnostic}
\vspace{3pt}
\centering
\setlength{\tabcolsep}{5pt}
\renewcommand{\arraystretch}{1.08}
\small
\begin{tabular}{l r r r}
\toprule
\textbf{Dataset}
  & \textbf{Shape $r$ $\uparrow$}
  & \textbf{Offset $\bar{\delta}$}
  & \textbf{Scale needed} \\
\midrule
\multicolumn{4}{l}{\textsc{In-Sample Test}} \\[2pt]
ISU\_ILCC        & 0.944 & $+0.161$ & 2.448 \\
NASA Rand        & 0.970 & $+0.218$ & 1.286 \\
TJU          & 0.976 & $+0.048$ & 1.267 \\
MATR         & \textbf{0.768} & $-0.100$ & 1.709 \\
XJTU-Partial     & 0.869 & $-0.202$ & 1.419 \\
\rowcolor{rowgray}
\textit{Mean}    & 0.905 & $+0.025$ & 1.626 \\
\midrule
\multicolumn{4}{l}{\textsc{Zero-Shot Test}} \\[2pt]
NASA PCoE        & \textbf{0.700} & $-0.206$ & 0.813 \\
Stanford-EV      & 0.805 & $-0.620$ & 0.422 \\
HUST   & \textbf{0.798} & $-0.056$ & 1.920 \\
XJTU-Full        & 0.874 & $-0.090$ & 3.069 \\
evTOL$^\dagger$  & \textbf{0.064} & $+0.303$ & 0.856 \\
\rowcolor{rowgray}
\textit{Mean (excl.\ evTOL)} & \textbf{0.794} & $-0.243$ & 1.556 \\
\midrule
\rowcolor{rowgray}
\textit{All (excl.\ evTOL)} & 0.856 & $-0.094$ & 1.595 \\
\bottomrule
\end{tabular}
\vspace{3pt}
\begin{minipage}{\columnwidth}
  \footnotesize
  $^\dagger$evTOL shape $r \approx 0$ reflects non-monotone capacity
  from aircraft partial-discharge profiles, not a representation failure
  ($|\rho|_{\mathrm{PC1}} = 0.997$).
\end{minipage}
\end{table}

The errors corresponding to the test and validation datasets are analysed to determine what factors are being under-represented in the jointly trained SOH model. In this study, we compute Pearson shape correlation $r$ between the centred predicted and true SOH trajectories, alongside the signed mean offset
$\bar{\delta} = \overline{\hat{SOH}} - \bar{SOH}$ and the scale ratio
$\sigma_{SOH} / \sigma_{\hat{SOH}}$, and average them per battery dataset. Table~\ref{tab:shape_diagnostic} reports the full shape diagnostic. 
The central finding is that the model's failure is not structural but
affine: shape correlations $r \geq 0.7$ on all non-evTOL datasets
indicate the trajectory shape is correctly learned from the frozen recurrent state trajectory alone. What varies across batteries is the
\emph{scale} and \emph{offset} of the predicted fade curve, not its
temporal structure.

The scale ratio $\sigma_{SOH} / \sigma_{\hat{SOH}}$ reveals two qualitatively
distinct failure modes.
\textit{Under-prediction of fade amplitude}
($\sigma_{SOH} / \sigma_{\hat{SOH}} > 1$): ISU\_ILCC ($2.45$),
XJTU-Full ($3.07$), 
HUST ($1.92$), and NASA~Rand ($1.28$).
In these datasets the model predicts a trajectory that fades too slowly.
These are predominantly partial-cycling datasets where the
effective depth of discharge is low and the absolute SOH change per FEC
is small, making the scale particularly hard to estimate through joint SOH head training.
\textit{Over-prediction of fade amplitude}
($\sigma_{SOH} / \sigma_{\hat{SOH}} < 1$): Stanford-EV ($0.42$),
NASA~PCoE ($0.81$).
Stanford-EV uses dynamic EV charging and discharging profiles with
silicon-graphite anodes, making it chemically distinct from all
training datasets. The model's PID-R features encode a faster-than-true
fade rate for this chemistry, producing an over-compressed trajectory.
Both failure modes are corrected by the per-dataset MLP, which learns
a dataset-specific nonlinear mapping that implicitly absorbs the
scale correction.

\subsection{Per-Battery GP Baseline}
\label{app:gp_baseline}

We additionally evaluated a per-battery Gaussian Process baseline with
a Mat\'ern-3/2 kernel, fitting one GP per battery on the first $k$ observed
SOH values and predicting the remainder.
On test datasets with fast degradation and chemically distinct profiles
(Stanford-EV: 33\%, NASA~PCoE: 18--28\%), the GP achieves competitive
results, confirming that the PC1 feature encodes a useful degradation
index for these batteries.
However, the GP fails systematically on slow-degrading datasets:
ISU\_ILCC at $k=5$ achieves $338\%$ MAPE because the first $k$
observations span negligible SOH range, leaving the scale parameter
entirely unconstrained.
This reveals a fundamental limitation of 1-D input GPs for
slow-fade batteries: the GP finds it hard to distinguish whether the battery has not degraded much in $k$ cycles or its degradation is invariant to the principal component, when both
produce flat training targets.
The per-dataset MLP sidesteps this by using the full $4d_h$-dimensional
PID feature vector, which provides richer degradation-rate cues than
the 1-D normalised PC1 score.

\subsection{Full Per-Dataset MLP Sweep}
\label{app:per_dataset_mlp}
 
\begin{table*}[t]
\caption{%
  Per-dataset MLP SOH head: MAPE (\%) and MAE across training
  fractions 10--50\% (mean over 3 random splits; std in parentheses on second line).evTOL and UMICH excluded.
  \textbf{Bold}: best result per dataset across fractions.
}
\label{tab:mlp_full_sweep}
\vspace{3pt}
\centering
\setlength{\tabcolsep}{4pt}
\renewcommand{\arraystretch}{1.0}
\resizebox{\textwidth}{!}{%
\small
\begin{tabular}{l
    cc cc cc cc cc}
\toprule
& \multicolumn{2}{c}{\textbf{10\%}}
& \multicolumn{2}{c}{\textbf{20\%}}
& \multicolumn{2}{c}{\textbf{30\%}}
& \multicolumn{2}{c}{\textbf{40\%}}
& \multicolumn{2}{c}{\textbf{50\%}} \\
\cmidrule(lr){2-3}\cmidrule(lr){4-5}\cmidrule(lr){6-7}
\cmidrule(lr){8-9}\cmidrule(lr){10-11}
\textbf{Dataset}
  & MAPE & MAE
  & MAPE & MAE
  & MAPE & MAE
  & MAPE & MAE
  & MAPE & MAE \\
\midrule
 
Stanford-EV
  & \makecell{6.2 \\ \scriptsize{($\pm$0.9)}}
  & \makecell{0.047 \\ \scriptsize{($\pm$.007)}}
  & \makecell{8.8 \\ \scriptsize{($\pm$3.1)}}
  & \makecell{0.065 \\ \scriptsize{($\pm$.018)}}
  & \makecell{\textbf{5.5} \\ \scriptsize{($\pm$0.6)}}
  & \makecell{\textbf{0.043} \\ \scriptsize{($\pm$.004)}}
  & \makecell{7.5 \\ \scriptsize{($\pm$1.6)}}
  & \makecell{0.058 \\ \scriptsize{($\pm$.014)}}
  & \makecell{6.0 \\ \scriptsize{($\pm$0.1)}}
  & \makecell{0.045 \\ \scriptsize{($\pm$.001)}} \\[2pt]
 
MATR
  & \makecell{32.4 \\ \scriptsize{($\pm$2.1)}}
  & \makecell{0.109 \\ \scriptsize{($\pm$.015)}}
  & \makecell{27.3 \\ \scriptsize{($\pm$2.1)}}
  & \makecell{0.086 \\ \scriptsize{($\pm$.008)}}
  & \makecell{21.6 \\ \scriptsize{($\pm$1.7)}}
  & \makecell{0.076 \\ \scriptsize{($\pm$.008)}}
  & \makecell{\textbf{20.9} \\ \scriptsize{($\pm$1.5)}}
  & \makecell{\textbf{0.070} \\ \scriptsize{($\pm$.007)}}
  & \makecell{20.3 \\ \scriptsize{($\pm$2.7)}}
  & \makecell{0.066 \\ \scriptsize{($\pm$.006)}} \\[2pt]
 
HUST
  & \makecell{32.6 \\ \scriptsize{($\pm$3.4)}}
  & \makecell{0.087 \\ \scriptsize{($\pm$.010)}}
  & \makecell{32.2 \\ \scriptsize{($\pm$4.7)}}
  & \makecell{0.078 \\ \scriptsize{($\pm$.002)}}
  & \makecell{30.9 \\ \scriptsize{($\pm$2.6)}}
  & \makecell{\textbf{0.074} \\ \scriptsize{($\pm$.001)}}
  & \makecell{32.3 \\ \scriptsize{($\pm$2.7)}}
  & \makecell{0.070 \\ \scriptsize{($\pm$.003)}}
  & \makecell{\textbf{31.4} \\ \scriptsize{($\pm$3.5)}}
  & \makecell{0.076 \\ \scriptsize{($\pm$.005)}} \\[2pt]
 
XJTU-Full
  & \makecell{33.3 \\ \scriptsize{($\pm$3.4)}}
  & \makecell{0.093 \\ \scriptsize{($\pm$.004)}}
  & \makecell{33.8 \\ \scriptsize{($\pm$4.2)}}
  & \makecell{0.074 \\ \scriptsize{($\pm$.001)}}
  & \makecell{32.1 \\ \scriptsize{($\pm$1.9)}}
  & \makecell{0.070 \\ \scriptsize{($\pm$.003)}}
  & \makecell{25.9 \\ \scriptsize{($\pm$1.6)}}
  & \makecell{0.065 \\ \scriptsize{($\pm$.005)}}
  & \makecell{\textbf{24.4} \\ \scriptsize{($\pm$1.3)}}
  & \makecell{\textbf{0.061} \\ \scriptsize{($\pm$.002)}} \\[2pt]
 
XJTU-Partial
  & \makecell{43.3 \\ \scriptsize{($\pm$6.3)}}
  & \makecell{0.162 \\ \scriptsize{($\pm$.073)}}
  & \makecell{36.8 \\ \scriptsize{($\pm$11.2)}}
  & \makecell{0.159 \\ \scriptsize{($\pm$.083)}}
  & \makecell{39.0 \\ \scriptsize{($\pm$9.5)}}
  & \makecell{0.158 \\ \scriptsize{($\pm$.092)}}
  & \makecell{37.2 \\ \scriptsize{($\pm$6.0)}}
  & \makecell{0.090 \\ \scriptsize{($\pm$.007)}}
  & \makecell{\textbf{24.6} \\ \scriptsize{($\pm$1.1)}}
  & \makecell{\textbf{0.092} \\ \scriptsize{($\pm$.006)}} \\[2pt]
 
TJU
  & \makecell{58.8 \\ \scriptsize{($\pm$6.6)}}
  & \makecell{0.140 \\ \scriptsize{($\pm$.005)}}
  & \makecell{\textbf{50.4} \\ \scriptsize{($\pm$3.5)}}
  & \makecell{\textbf{0.117} \\ \scriptsize{($\pm$.001)}}
  & \makecell{51.3 \\ \scriptsize{($\pm$0.7)}}
  & \makecell{0.128 \\ \scriptsize{($\pm$.003)}}
  & \makecell{53.5 \\ \scriptsize{($\pm$1.0)}}
  & \makecell{0.131 \\ \scriptsize{($\pm$.006)}}
  & \makecell{57.8 \\ \scriptsize{($\pm$8.3)}}
  & \makecell{0.124 \\ \scriptsize{($\pm$.004)}} \\[2pt]
 
ISU\_ILCC
  & \makecell{94.0 \\ \scriptsize{($\pm$13.7)}}
  & \makecell{0.402 \\ \scriptsize{($\pm$.014)}}
  & \makecell{89.2 \\ \scriptsize{($\pm$11.0)}}
  & \makecell{0.378 \\ \scriptsize{($\pm$.011)}}
  & \makecell{71.4 \\ \scriptsize{($\pm$1.7)}}
  & \makecell{0.355 \\ \scriptsize{($\pm$.011)}}
  & \makecell{73.4 \\ \scriptsize{($\pm$2.7)}}
  & \makecell{0.355 \\ \scriptsize{($\pm$.005)}}
  & \makecell{\textbf{70.5} \\ \scriptsize{($\pm$2.3)}}
  & \makecell{\textbf{0.340} \\ \scriptsize{($\pm$.003)}} \\[2pt]
 
NASA Rand
  & \makecell{86.0 \\ \scriptsize{($\pm$11.8)}}
  & \makecell{0.690 \\ \scriptsize{($\pm$.072)}}
  & \makecell{91.1 \\ \scriptsize{($\pm$24.5)}}
  & \makecell{0.657 \\ \scriptsize{($\pm$.120)}}
  & \makecell{\textbf{76.6} \\ \scriptsize{($\pm$13.2)}}
  & \makecell{\textbf{0.637} \\ \scriptsize{($\pm$.025)}}
  & \makecell{79.0 \\ \scriptsize{($\pm$32.5)}}
  & \makecell{0.678 \\ \scriptsize{($\pm$.240)}}
  & \makecell{96.4 \\ \scriptsize{($\pm$47.0)}}
  & \makecell{0.792 \\ \scriptsize{($\pm$.338)}} \\
 
\bottomrule
\end{tabular}}
\vspace{3pt}
\begin{minipage}{\textwidth}
  \footnotesize
  Stanford-EV saturates at $\sim$6\% from 10\%.
  XJTU-Full and MATR improve monotonically.
  HUST plateaus at $\sim$31\% (irreducible intra-dataset variability).
  NASA~Rand exhibits high variance from small pool and random
  discharge protocols.
\end{minipage}
\end{table*}

Here, we evaluate how much of the residual affine error can be corrected by training a per-dataset MLP head on fractions of in-domain batteries, using the frozen PID-R features as input. We consider 5 training fractions regimes each using one of 5 fractions between 10-50 \% of the dataset for training. Table~\ref{tab:mlp_full_sweep} reports MAPE and MAE across all training
fractions for all eight datasets.
Several patterns are worth noting.
Stanford-EV achieves $5$--$9\%$ MAPE across these fractions of training batteries
and barely improves with additional data, indicating that Stanford-EV
batteries are highly homogeneous in their degradation behaviour and a
small number of training batteries suffice to characterise the distribution.
In contrast, XJTU-Full improves monotonically from $33\%$ to $24\%$ as
training fraction increases from $10\%$ to $50\%$, suggesting the
model has not yet saturated its capacity to learn from additional batteries
— more data would likely continue to improve results.
HUST plateaus at $31$--$33\%$ regardless of training fraction,
indicating that the $32\%$ floor is irreducible under the current
representation: the remaining error reflects genuine battery-to-battery
capacity variability within the HUST dataset that is not encoded in the
voltage-forecasting hidden states.
TJU shows a non-monotone trend, peaking at $20\%$ training fraction
before degrading slightly at higher fractions, suggestive of mild
overfitting in the MLP head when more heterogeneous training batteries
are included.
NASA~Rand exhibits high variance (up to $\pm 47\%$ at $50\%$) due to the
small dataset size and highly non-stationary random discharge protocols; reliable conclusions about this dataset
require more batteries.
The ISU\_ILCC batteries undergo deep degradation (much below EOL) and possibly have the largest intra dataset variability of the lot, leading to the high errors in the SOH modelling task. 

Further analysis of this could extend into trying different features from the recurrent states as input to such per dataset SOH heads. 

\printbibliography

\end{document}